\documentclass[acmsmall,nonacm]{acmart}

\usepackage{amsmath,amsfonts}
\usepackage{array}
\usepackage{booktabs}
\usepackage{placeins}

\graphicspath{{figures/}}

\copyrightyear{2026}
\acmYear{2026}

\renewcommand\footnotetextcopyrightpermission[1]{}
\begin{document}

\title{Natural Language Interfaces for Databases: What Changes for SQL-Literate Users?}

\author{Panos Ipeirotis}
\email{panos@nyu.edu}
\affiliation{%
  \institution{Stern School of Business, New York University}
  \department{Department of Technology, Operations, and Statistics}
  \city{New York}
  \state{NY}
  \postcode{10012}
  \country{USA}
}

\author{Haotian Zheng}
\email{hz2687@nyu.edu}
\affiliation{%
  \institution{Tandon School of Engineering, New York University}
  \department{Electrical and Computer Engineering}
  \city{New York}
  \state{NY}
  \postcode{10012}
  \country{USA}
}

\begin{abstract}
Natural Language Interfaces for Databases (NLIDBs) let users query data in everyday language instead of SQL, and modern systems translate most questions accurately. Does an NLIDB remove the work of querying, or move it? We compared SQL-LLM, a GPT-4o-backed NLIDB, with Snowflake, a manual-SQL platform. Twenty SQL-literate professionals and graduate students, ten per interface, each completed 12 BIRD benchmark tasks with hints naming target columns and operators. SQL-LLM cut completion time per query by 31\%. It did not improve accuracy: SQL-LLM users answered 46\% of queries correctly versus 64\% for Snowflake, a gap this sample cannot confirm ($p = 0.08$) that narrows to 14 points without one session where the system timed out. Think-aloud audio from 11 participants (4 SQL-LLM, 7 Snowflake) showed how the work shifted: SQL-LLM users left initial schema navigation to the model, mainly on easier tasks, and focused on checking the generated SQL, while Snowflake users explored the schema and wrote queries by hand. Coded behavior rates did not differ detectably between arms, and positive remarks during a task did not reliably predict a correct answer. The evidence fits a reallocation of querying work, not its removal: users still had to check the generated SQL.
\end{abstract}

%


\maketitle

\section{Introduction}\label{sec:intro}

Natural Language Interfaces for Databases (NLIDBs) promise a simple bargain: type a question, get an answer, skip the SQL. The bargain responds to a long-standing complaint about how hard database systems are to use~\cite{Jagadish2007}. The translation NLIDBs perform, from a natural-language question to SQL (Text-to-SQL), is now technically credible, with recent large language models (LLMs) reaching roughly 85\% accuracy on benchmarks like Spider~\cite{pourreza2023gpt4}. The harder question is what users do with these systems: how they verify generated queries, where they get stuck, and whether the interface earns enough trust to displace a manual-SQL workflow~\cite{ning2023empirical, kochedykov-etal-2023-conversing}. One controlled study found that interactive interfaces for repairing generated SQL gave users no significant gain in accuracy or completion time over editing the query directly~\cite{ning2023empirical}, but whether a modern LLM-based NLIDB beats a hand-written SQL workflow for SQL-literate users has not been tested head-to-head.

We test that question in a controlled, mixed-method study comparing SQL-LLM, a deployed NLIDB backed by GPT-4o, against a manual SQL workflow in Snowflake's Snowsight editor (referred to simply as ``Snowflake'' below). Twenty SQL-literate professionals and graduate students each completed 12 tasks from the BIRD Text-to-SQL benchmark~\cite{li2024can} on three relational schemas, recorded with screen capture, think-aloud protocols, and timing logs. Both groups queried the same Snowflake warehouse and execution engine (Section~\ref{sec:databases}), so only the interface differed: natural-language prompting, model-proposed tables, and an editable generated query versus writing SQL from scratch.

SQL-LLM was substantially faster: it cut completion time per query by 31\% on the log scale ($p = 0.03$), or about $212$ s per query on the raw scale ($p = 0.04$--$0.07$; Section~\ref{sec:overall}). That speed did not translate into higher accuracy. Graded against the BIRD gold answers, SQL-LLM users answered 46\% of queries correctly versus 64\% for Snowflake, an 18-point gap favoring manual SQL ($p = 0.08$ at $N = 20$) that narrows to 14 points after excluding one SQL-LLM session where the system repeatedly timed out (Section~\ref{sec:results}). Because faster completion offset lower accuracy, both groups produced roughly the same number of correct answers per hour (4.6 for SQL-LLM versus 4.2 for Snowflake). The interface changed how participants worked, not how many right answers they delivered per hour.

Think-aloud recordings from 11 participants (4 SQL-LLM, 7 Snowflake) show how that work shifted. SQL-LLM users let the model handle initial schema navigation, especially on easier tasks, and spent their time checking whether the generated SQL matched their intent. Snowflake users instead explored the schema upfront and wrote queries by hand. Neither group showed a detectable difference in per-minute rates of frustration, confusion, schema checks, hesitation, reformulation, or positive confirmations. Just as important, participants' positive remarks during a task did not reliably predict whether their final answer was right in either interface. Together, these patterns point to a reallocation of querying effort rather than its elimination.

Three design limits qualify that conclusion, and we state them upfront. First, our sample of 10 participants per arm can detect only medium-to-large effects, leaving the 18-point accuracy gap and the per-minute behavioral differences statistically inconclusive. Second, only 11 of 20 sessions had audio clear enough for behavioral coding (4 SQL-LLM, 7 Snowflake); schema checks and confusion point in the same direction across nearly all balanced 4-versus-4 subsamples, but we cannot rule out selection effects from who spoke aloud. Third, every task included a hint naming target columns and operators, sometimes in near-SQL notation (for example, glossing ``most orders'' as \texttt{MAX(COUNT(order\_id))}). Because those hints did part of the schema lookup for both groups, our results reflect heavily disambiguated tasks rather than open-ended questions (Section~\ref{subsec:taskdesign}).

Even with those caveats, shifting work from query writing to query verification has clear implications for analytics teams. NLIDBs that promise to ``democratize data access''~\cite{hong2025next} still leave users checking whether the generated query matches the question they asked~\cite{tian2024sqlucid}. Our participants knew SQL and could inspect the model's queries directly; how users without SQL training handle that verification step remains an open question. Among SQL-literate users, natural-language input does not remove the need to verify the query.

\paragraph*{Contributions.}
\begin{itemize}
    \item \textbf{Reallocation of querying work.} A behavioral analysis of think-aloud transcripts and screen recordings showing how effort shifts between interfaces: manual SQL users invest effort upfront in schema exploration and query writing, SQL-LLM users invest effort downstream in verifying generated SQL, and in neither group do positive remarks reliably indicate a correct answer.

    \item \textbf{Head-to-head comparison.} A controlled user study comparing a deployed GPT-4o NLIDB against a manual SQL editor on the same Snowflake warehouse (20 SQL-literate participants, 12 BIRD tasks each, 11 double-coded think-aloud sessions).

    \item \textbf{Speed, accuracy, and throughput analysis.} A mixed-effects and sensitivity analysis showing a 31\% per-query speed gain alongside a non-significant 18-point accuracy gap ($p = 0.08$) and near-equal correct answers per hour.

    \item \textbf{Adoption hypotheses for analytics teams.} Practical hypotheses on review skills, schema governance, query visibility, and staged rollout when introducing NLIDBs to SQL-literate teams.
\end{itemize}

Section~\ref{sec:related} positions the study against prior comparisons of NLIDBs with traditional query interfaces. Section~\ref{sec:methodology} describes the setup and study design; Section~\ref{sec:results} reports the results, with the qualitative analysis in Section~\ref{sec:qualitative}. Section~\ref{sec:discussion} discusses implications, Section~\ref{sec:futurework} outlines future work, and Section~\ref{sec:conclusion} concludes.

\section{Related Work}
\label{sec:related}

Benchmark accuracy for NLIDBs has climbed high on standard benchmarks, though it still trails skilled human annotators on harder benchmarks such as BIRD~\cite{li2024can}; the open question is whether that accuracy helps real users. The empirical user studies below converge on one unresolved question: once an NLIDB is accurate enough to deploy, does it reduce the user's remaining effort or merely relocate it? Prior work measured accuracy, speed, and preference across ways of handling NLIDB errors, and one finding sets up the tension this study takes up: interactive repair interfaces gave no significant gain in accuracy or speed over editing the generated SQL by hand~\cite{ning2023empirical}.

\subsection{From Benchmark Accuracy to the Verification Burden}

Querying a database in plain language is an old promise, answering a long-standing complaint about how hard database systems are to use~\cite{Jagadish2007}; what changed recently is accuracy. On large-scale benchmarks like Spider~\cite{yu2018spider}, research shifted toward deep neural models~\cite{kochedykov-etal-2023-conversing}, and GPT-4-based methods have pushed accuracy further, reporting about 85\% execution accuracy on the Spider test set~\cite{pourreza2023gpt4}.

High accuracy does not settle whether the system helps the person using it; the gap that remains is one of understanding and validating the generated query, on which trust depends~\cite{tian2024sqlucid}. A residual error rate of roughly 15\% on Spider~\cite{ning2023empirical} matters in business-critical domains, where incorrect data drives flawed decisions. Users respond by demanding near-perfect precision~\cite{kochedykov-etal-2023-conversing}, yet struggle to find and repair errors in system-generated SQL, especially semantically complex ones~\cite{ning2023empirical}. That struggle predates LLMs: from exam analyses~\cite{brass2006semantic} to student corpora of over 33,000 queries~\cite{taipalus2018errors}, studies of hand-written SQL show that students learning SQL routinely write queries that are syntactically legal but semantically wrong (contradictory conditions, missing join predicates), which the DBMS executes without warning.

The problem sharpens when a system behaves as a ``black box,'' returning an answer with no way to check it. The model leaves a residual error rate; opacity makes those errors impossible to inspect, so users cannot calibrate their trust. Checking the generated SQL against the question asked is the one act that lets a user detect and manage the errors, though it does not shrink the error rate itself. That checking work is the verification burden, and it frames the design response the rest of this section traces: make the generated SQL visible so the user can verify it. An early vision proposed the complementary move, rendering a query back into natural language so the user could confirm it captured their intent~\cite{Simitsis2009}. Recent research follows this line, moving the user from accepting opaque output to inspecting it.

\subsection{Comparing NLIDBs with Traditional Query Interfaces}

Direct head-to-head evidence on natural-language querying versus manual SQL is old, and none of it concerns users who already knew SQL. The comparisons of the 1970s and 1980s, in the laboratory~\cite{Shneiderman1978} and in the field~\cite{Jarke1985}, trained novices in SQL for the study and compared their SQL with queries posed in English; the field study found SQL better on most measures, though natural language took less effort to use~\cite{Jarke1985}. A third, a paper-and-pencil test in which no system ran, found no significant difference in accuracy, with natural-language users faster; the point estimates leaned toward SQL, with 53\% of queries essentially correct against 45\% in English~\cite{Vassiliou1983}. Our accuracy gap points the same way and is larger (18 percentage points). What changed between a 1983 paper-and-pencil test and a modern LLM interface is how failures look to the user. A paper-and-pencil test returns no output at all, whereas an LLM almost always returns SQL that executes cleanly and produces a plausible table. Most wrong answers in our study failed silently in just this way (Section~\ref{sec:results}); visible failures such as timeouts were the exception. Because tasks, populations, and interfaces also differ across the two studies, silent failure is an explanatory hypothesis rather than a tested mechanism. How a modern LLM-based interface performs for users who already know SQL has not previously been tested head-to-head.

The most rigorous recent user study shifts the question from efficiency to verification. Ning et al.~\cite{ning2023empirical} tested interactive error-handling interfaces (a step-by-step decomposition view, a graph visualization, and a conversational dialogue) against a no-support baseline in which users directly edited the model-generated SQL, and found no significant overall difference in task completion time or accuracy: the added interfaces bought no measurable gain over editing the generated query by hand, although participants preferred the step-by-step explanation interface and non-experts found it more useful. That study did not compare an NLIDB against a hand-written SQL workflow; ours does, for SQL-literate users.

The benefit prior work does report sits elsewhere, with users who lack SQL. NaLIR let people with little knowledge of SQL formulate complex queries involving joins and aggregations, completing 88 of 98 tasks against 56 with a faceted search interface~\cite{Li2014, Li2016}, widening the range of questions a non-technical user can ask. That result casts Text-to-SQL as a complement, not a replacement, most valuable to users without SQL skills or for queries hard to express otherwise. Our study does not test non-technical users and makes no claim about them; it asks what a modern NLIDB does for the SQL-literate population, for which direct evidence has been missing.

\subsection{Lowering the Verification Burden: Interactive and Explainable Systems}

Interactive and explainable NLIDBs are the field's answer to the verification burden: each makes its logic transparent and hands the user something to verify. Their evaluations measure error discovery, repair success, confidence, and SQL comprehension, but none asks how a user's effort redistributes between construction and verification when the interface changes.

\paragraph{Decomposition and Step-by-Step Explanations.} One approach breaks a complex query into simpler steps, each explained in natural language. The DIY system, designed for users without SQL expertise, splits a query into sub-queries and shows each one's result on a small sample database; in Narechania et al.'s exploratory study, participants checked queries by inspecting these steps and editing the sample data, though some struggled to follow the explanations of joins~\cite{Narechania2021}. STEPS extends this line with editable step-by-step natural-language explanations of the generated query, evaluated in a 24-participant user study by the group that later built SQLucid~\cite{tian2023steps}. The step-by-step view lowers the cost of verification without removing it: the user still reads each step's output and judges whether it is right.

\paragraph{Visual Query Representations.} Another strategy visualizes query structure. QueryVis~\cite{Leventidis2020} and SQLVis~\cite{Miedema2021} render tables, joins, and filters as diagrams, and Leventidis et al. found that QueryVis diagrams let users interpret queries about 20\% faster than SQL text, with weaker evidence of fewer errors~\cite{Leventidis2020}. The diagram is a cheaper route to the same verification: it speeds reading of a query the user must still confirm matches their intent.

\paragraph{Conversational Clarification and Direct Manipulation.} A third direction opens a dialogue about the parts of a generated query the system is unsure of. MISP~\cite{Yao2019} and DialSQL~\cite{gur2018dialsql} ask follow-up questions: MISP asks the user to confirm a predicted column or condition, and DialSQL offers alternatives to choose from (e.g., ``sales amount'' or ``number of sales''). In Ning et al.'s study the dialogue was the least preferred mechanism: it often introduced new errors, and users quickly grew frustrated~\cite{ning2023empirical}. The dialogue moves verification upstream into the user's answers, yet still leaves them to judge whether the resulting query is correct.

\subsection{Empirical Evidence on User Performance and Confidence}

When these systems are tested on real users, the payoff tracks how well the design supports verification, not how clever the translation is. The Ning et al. study behind the null result above~\cite{ning2023empirical} attributed that null mainly to its harder, state-of-the-art error set, and noted that the tools gave users little help validating their own repairs. Transparency helps only when it makes the error checkable.

Combining modalities helps more. SQLucid pairs natural-language explanations with visual highlights and direct editability, letting users see how their input maps to the resulting SQL~\cite{tian2024sqlucid}. In a 30-participant study it raised task-completion accuracy to 85\%, against 56\% for MISP and 67\% for DIY, narrowed the novice-expert gap, and lifted confidence (6.4/7 vs.\ 3.8/7 for MISP and 5.3/7 for DIY). Users found the system's mistakes easy to correct, the same lesson from the other side: the design has to support the user's verification burden, not just the model's translation.

Our study takes up that open question for SQL-literate users directly: a deployed GPT-4o-backed NLIDB against a conventional SQL editor on the same task set, both arms backed by the same Snowflake warehouse. Section~\ref{sec:discussion} returns to where our results echo Ning et al.~\cite{ning2023empirical} and where they diverge.
The behavioral pattern we document complements the SQLucid~\cite{tian2024sqlucid} finding that interactive explainability raises confidence: the observable work of querying shifts from schema and query construction toward checking generated SQL, and both results locate NLIDB usability in the user's verification burden, not the model's translation alone.

\section{Methodology}\label{sec:methodology}

We compare SQL-LLM and Snowflake in a between-subjects study. Twenty SQL-literate participants were randomly assigned to one of the two interface arms, ten per arm, and every participant attempted the same 12 tasks, drawn from the BIRD benchmark, on three relational schemas. The outcome measures are per-query completion time and answer accuracy graded against the BIRD gold answers, both collected for all 20 participants; behavioral coding of the think-aloud commentary covers the 11 sessions with codable audio. We first describe what both arms hold constant and what the SQL-LLM arm adds (Experimental Setup), then who took part, what they were asked to do, and what we measured (Study Design).

\subsection{Experimental Setup}

The two arms share the same data, schemas, and execution engine, so any difference in completion time or accuracy comes from the interface itself. Our comparison tests the complete SQL-LLM package (natural-language prompting, table suggestions, query generation and validation, an editable SQL pane, and follow-up chat) against a manual SQL editor. Because the study manipulates that whole package at once, it cannot isolate the effect of any single feature; later claims about individual features are therefore design hypotheses rather than causal estimates. We first describe what both arms share (databases, tasks, and difficulty levels) and then describe the SQL-LLM interface.

\subsubsection{Databases Used}
\label{sec:databases}

The study used three relational databases from the BIRD dataset, each from a distinct domain. \textbf{Books} is a bibliographic schema of authors, books, publishers, and publication years, supporting queries on authorship, title filtering, and publication metrics. \textbf{Mondial Geo} is a geographic schema of countries, cities, populations, and regions that invites multi-table joins and aggregations, such as comparing populations across regions or filtering cities by country. \textbf{Legislator} covers U.S. legislator names, terms served, state affiliations, and demographics, supporting queries on service history and party affiliation.

All three were ingested as native tables in a single Snowflake database and queried through one consistently configured virtual warehouse (Snowflake's compute unit): the Snowflake arm through the standard Snowsight web UI, and the SQL-LLM arm through the SQL-LLM editor, which generated SQL against the same warehouse. Both arms therefore queried the same data over the same execution engine and compute configuration, which removes query-engine performance as a confound.

\subsubsection{Task Selection}

The study tasks come from BIRD (a BIg Bench for LaRge-scale Database Grounded Text-to-SQLs)~\cite{li2024can}, a dataset of realistic natural-language questions mapped to SQL across multiple domains. We selected 12 tasks (4 per database) spanning common business-intelligence patterns (entity retrieval, aggregated reporting, nested-condition queries) and varying across three difficulty levels (Easy, Medium, and Hard; Section~\ref{subsec:difficulty}).

\subsubsection{Difficulty Categorization}
\label{subsec:difficulty}

To relate behavior to query difficulty, we categorized the 12 tasks into three levels:

\begin{itemize}
    \item \textbf{Easy Level}: single-table selection with basic filters and projections.
    \item \textbf{Medium Level}: multi-table joins with aggregations or group-by operations.
    \item \textbf{Hard Level}: nested subqueries, multiple joins, and conditional filtering logic.
\end{itemize}

The set is balanced four tasks to a level; the per-database Easy/Medium/Hard split is Books 1/2/1, Mondial Geo 1/1/2, and Legislator 2/1/1. Levels reflect schema-comprehension demands and structural complexity, our own structural heuristics rather than BIRD's published difficulty labels.

Difficulty enters the linear mixed-effects model as a covariate (Section~\ref{sec:overall}) to account for differences across the 12 tasks. Because our tasks come from BIRD's public training split, which does not include per-question difficulty labels (unlike BIRD's development split), we checked our Easy, Medium, and Hard groupings against an objective structural complexity score computed from each task's gold SQL. The score counts referenced tables and aggregation functions, adds one point each for \texttt{GROUP BY}, \texttt{HAVING}, \texttt{ORDER BY}, and \texttt{DISTINCT} clauses, and gives double weight to nested subqueries and set operations. Mean structural complexity rises across the three levels ($2.00$ for Easy, $3.25$ for Medium, and $5.50$ for Hard; Spearman $\rho = 0.69$). Individual tasks still overlap at the boundaries: scores range from 1 to 3 for Easy, 1 to 5 for Medium, and 3 to 8 for Hard, so the most complex Easy task (L-Q4) ties the simplest Hard task (M-Q2). The scoring script (\texttt{difficulty\_structural\_check.py}) and output table (\texttt{difficulty\_structural\_complexity.csv}) are in the replication archive.

\subsubsection{SQL-LLM System Overview}
\label{sec:sqlllm-ui}
SQL-LLM is a commercially deployed NLIDB backed by OpenAI's GPT-4o (via the OpenAI API in spring 2025); we evaluate it under a pseudonym at the vendor's request. On top of the shared warehouse (Section~\ref{sec:databases}), it adds the interface bundle of natural-language prompting, model-proposed tables, an editable generated query, and re-run.

\begin{figure}[t]
    \centering
    \includegraphics[width=\linewidth]{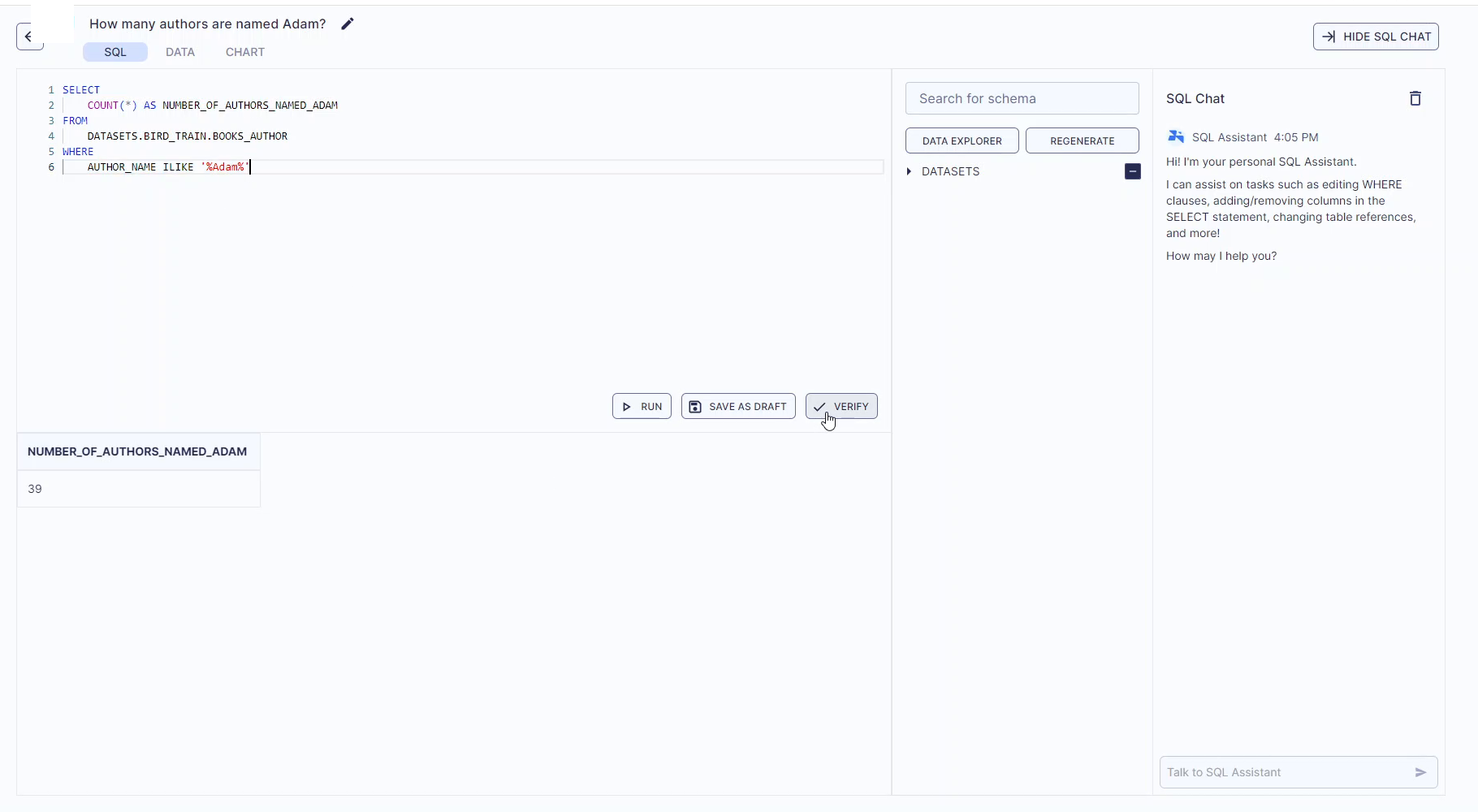}
    \caption{The SQL-LLM editor. The natural-language question (top), the generated SQL (center, editable), the result table (bottom), and the conversational SQL Chat pane (right) were all visible at once; the collapsible \textsc{Datasets} panel, where users picked from the tables SQL-LLM proposed, is not shown. \textsc{Run}, \textsc{Regenerate}, and \textsc{Verify} were the three primary actions; participants could also edit the SQL directly or issue follow-up instructions in the chat pane.}
    \label{fig:sql_llm_ui}
\end{figure}

Between the user and the model sits a layer that turns each natural-language question into validated SQL: it serializes the active database schema, assembles a few-shot Text-to-SQL prompt with curated exemplars for that database, forwards the question to GPT-4o, and checks that the generated SQL runs against the active schema; queries that fail are rejected rather than displayed. The exemplars are fixed per database, not retrieved per question: each study database carries a curated set of 5--10 question/SQL pairs drawn from the BIRD and Spider training splits, attached at deployment, and the full per-database set accompanies every question on that database. Validated SQL appears in an editable editor (Figure~\ref{fig:sql_llm_ui}) and executes against that same Snowflake warehouse when the user presses \textsc{Run}.

None of the 12 study tasks appeared in any database's few-shot prompt: we screened the exemplars against the study tasks and excluded any match, so the model never saw a test question as an in-context example. That screen only covers the prompt, however; because BIRD and Spider are public benchmarks, GPT-4o may have seen their schemas or questions during pretraining. Ruling out pretraining exposure requires privately authored tasks on custom schemas (Section~\ref{sec:futurework}).

The system ran without human intervention during the study: the validator blocked unrunnable queries without rewriting them, and whatever SQL a participant submitted via \textsc{Run} executed verbatim on the shared warehouse. Two logging gaps limit how finely we can dissect those interactions. First, validator rejections were not logged separately, so time spent re-prompting after a rejected generation is included in total task completion time. Second, intermediate interaction traces (the model's initial SQL before user edits, regeneration clicks, and chat turns) were not recorded, so we infer repair behavior from the screen and think-aloud recordings rather than server logs.

\paragraph{Interface affordances.}
Figure~\ref{fig:sql_llm_ui} shows the SQL-LLM editor as participants saw it, with the natural-language question and the model's generated SQL side by side. The generated SQL was always visible in the editor pane, so participants could read and confirm it against their intent before running. The editor was a free-form SQL textarea: participants could modify the model's output (rename columns, add filters, fix joins) and re-execute, or write SQL from scratch. A \textsc{Regenerate} button re-issued the original question to the model for an alternate query, and a conversational \emph{SQL Chat} pane let participants issue follow-up natural-language instructions (e.g., ``add a filter for year > 2000'') that the model applied incrementally to the current query. Generation did not auto-execute: participants pressed \textsc{Run} to send the SQL to the warehouse, and results appeared as a table beneath the editor. A collapsible \textsc{Datasets} panel exposed the database, schema, table, and column hierarchy with type-to-filter search. Before each generation, SQL-LLM proposed in this panel the tables relevant to the question, and the user picked the ones the model should use, so the model, not the user, did the initial search of the schema. A \textsc{Verify} toggle let participants mark a result as accepted, a participant-side bookkeeping affordance that did not feed back into the model.

\subsection{Study Design}

\subsubsection{Participants}\label{subsec:participants}

We recruited 20 participants from Upwork, an online freelancing platform, screening for a work history as an analyst with SQL skills: working data analysts, business-intelligence professionals, and graduate students trained in databases. All reported at least basic SQL familiarity, and none had prior exposure to the SQL-LLM interface. Our findings therefore speak to SQL-literate users; we do not test, and do not claim, benefit for non-technical users who cannot read or write SQL.

All participants consented to be recorded. The protocol was approved by the New York University Institutional Review Board (IRB-FY2023-7595); in quotations and tables, participants appear only by neutral codes (P1--P10 for the SQL-LLM arm, P11--P20 for Snowflake), matching the public replication package. Each was paid a fixed \$100 via Upwork and could take breaks or withdraw without penalty. Payment was not contingent on accuracy; Section~\ref{subsec:limitations} revisits the resulting satisficing risk.

\subsubsection{Task Design}
\label{subsec:taskdesign}

Each participant attempted the same 12 tasks on their assigned interface: the tasks selected above, posed as natural-language questions and grouped by database. Two representative example prompts appear below:

\begin{quote}
\textbf{Database:} \texttt{books}
\begin{itemize}
    \item \textbf{B-Q1.} Which customer has made the most orders? Show his/her full name.\\
    \textit{Hint:} Most orders refers to \texttt{MAX(COUNT(order\_id))}; customer refers to \texttt{first\_name, last\_name}.\\
    \item \textbf{B-Q2.} How many authors are named Adam?\\
    \textit{Hint:} Authors named Adam refers to \texttt{author\_name LIKE 'Adam'}.\\
\end{itemize}
\end{quote}

Each task included a text hint naming the target columns and operators (for example, glossing ``most orders'' as \texttt{MAX(COUNT(order\_id))}). These hints clarified the question but also did part of the work that the two interfaces handle differently: SQL-LLM users could paste or paraphrase the hint in their prompt, while Snowflake users got the exact column names and operators without searching the schema. Because both groups benefited in different ways, our design cannot tell which group gained more overall. Two specific hints deserve note. On B-Q2, the hint's literal predicate \texttt{LIKE 'Adam'} conflicts with the gold query's prefix match, a discrepancy we examine in Section~\ref{sec:results}. On B-Q4, the hint used pseudo-SQL (\texttt{COUNT(\ldots WHERE \ldots)}) rather than valid SQL syntax, which Snowflake users had to translate by hand while SQL-LLM users could pass directly to the model. Our findings therefore apply to heavily disambiguated tasks rather than open-ended questions (Section~\ref{subsec:limitations}).

Participants recorded their screens with voice commentary while working, and recordings with usable audio were transcribed with Whisper for the qualitative analysis. That analysis codes reformulations as a behavioral measure, so the count must be comparable across arms. We count a reformulation symmetrically across arms: a substantively different submission of the same task, re-executed after the participant saw the previous result. In the SQL-LLM arm, that is a re-submitted changed natural-language prompt or an edited-and-re-executed returned query; in the Snowflake arm, an edited-and-re-run query or a semantically new attempt. An edit never re-executed does not count in either arm, and neither do trivial typo fixes or identical re-runs.

\subsubsection{Procedure}
\label{subsec:procedure}
We enrolled the first 20 applicants who passed this screen (Section~\ref{subsec:participants}) and randomly assigned them to two equal groups of ten, one per arm; each completed all 12 tasks. Participants used their own hardware with a standardized browser configuration. Every participant received the 12 tasks in the same listed order and could skip a task and return to it later. Each session had three phases:

\begin{enumerate}
    \item \textbf{Training} (15 minutes): participants explored the schemas of all three databases and completed two practice queries on the assigned system.
    \item \textbf{Testing} (1 hour and 45 minutes): participants executed the 12 tasks while screen and audio recordings captured their behavior.
    \item \textbf{Post-task debrief} (up to 30 minutes): participants answered an open-ended questionnaire on usability and perceived difficulty. We have no reliable record that a structured instrument (e.g., NASA-TLX~\cite{hart1988} or trust scales) was properly administered, and no data from one. The free-text responses are not systematically analyzed, and the behavioral claims rest on the think-aloud record, not the debrief.
\end{enumerate}

These durations were advisory: the remote, self-paced sessions let participants exceed an allotment without penalty, and many did. Summed on-task time across the 12 tasks had a median of roughly 100 minutes, but 10 of the 20 participants (3 SQL-LLM, 7 Snowflake) exceeded the nominal 105-minute testing allotment, the longest by more than a factor of two. Per-task completion times come from the screen recordings (Section~\ref{subsec:measures}) and are unaffected by these allotments.

\paragraph{Coverage of think-aloud audio.} All 20 participants completed the 12 tasks and submitted screen recordings, so all completion-time and accuracy analyses (Tables~\ref{tab:overall}, \ref{tab:results}, and~\ref{tab:ttest}) use the full 20-participant sample. However, only 11 participants (4 SQL-LLM, 7 Snowflake) produced think-aloud audio clear enough for behavioral coding; the other 9 had muted microphones, missing audio tracks, or commentary too sparse to code. Because we filtered for usable audio after participants used their assigned interface, and because one interface might naturally prompt less talking than the other, we cannot rule out selection bias in which 11 sessions were coded. The 11 coded participants did not differ significantly from the 9 uncoded participants in mean completion time (558 vs.\ 481 s, Welch $p = 0.49$) or accuracy (60\% vs.\ 49\%, $p = 0.33$), though a comparison of 11 against 9 has limited statistical power. In Section~\ref{sec:qualitative}, a check across all 35 balanced 4-versus-4 subsamples shows that per-minute differences in schema checks and confusion keep the same sign in 34 and 33 subsamples. All transcript-based analyses in Section~\ref{sec:qualitative} are restricted to these 11 sessions.

\subsubsection{Measures}
\label{subsec:measures}

Two primary outcomes cover all 20 participants: per-query completion time and answer accuracy. Completion time is the active on-task duration logged from each participant's screen recording, excluding breaks away from the desk. Because it measures total wall-clock time on the task, it bundles together thinking, typing, model generation, and query execution without separating them (Section~\ref{sec:sqlllm-ui}). Timing logs cover 238 of the 240 participant--task pairs; two times for Snowflake participant P17 (M-Q1 and M-Q2) were not recorded, though both submitted queries were captured and graded, leaving 120 graded submissions per arm.

To grade accuracy, we ran each participant's final submitted query against the BIRD database and marked it correct if its output matched the BIRD gold query's result set. Before finalizing grades, we audited all 12 BIRD gold queries against the exact prompts shown to participants, correcting one mismatched gold query (M-Q2) and flagging one hint--gold discrepancy (B-Q2; Section~\ref{sec:results} and Appendix~\ref{app:goldaudit}). As a secondary check, three independent LLM judges scored all 240 submissions against the gold queries; their majority vote agreed with execution grading on 87.5\% of submissions, with a third of the disagreements coming from the uncorrected M-Q2 gold query.

Behavioral measures such as frustration and reformulation counts are coded from the think-aloud transcripts and cover only the 11 sessions with codable audio, using the symmetric reformulation definition in Section~\ref{subsec:taskdesign}. Per-minute rates divide each count by the session's audited length, its duration checked against the screen recording, because some transcript clocks paused or reset; seven of the 11 sessions have transcript clocks that match the recording.

\paragraph{Analysis pipeline.} Several research assistants worked on this project over time, and earlier versions of the analysis relied in part on derived values recorded by hand. We rebuilt every analysis from scratch, starting from the raw study data (timing logs, submitted queries, transcripts, and the two coders' event files). The rebuild used AI coding agents, chiefly Claude Code, which wrote the grading, statistics, and figure scripts under the authors' direction, with OpenAI Codex reviewing the code changes. Every number in this paper is regenerated by those scripts from the raw data, and before submission we re-ran the full pipeline end to end, from a fresh download of the BIRD databases through the execution grades, and confirmed that it reproduces every reported value. The behavioral coding itself was done by the two human analysts hired on Upwork; the LLM coders in Appendix~\ref{app:llmcoders} serve only as a cross-check. Two later annotations used AI agents, each reading the full transcripts and recording verbatim evidence for every decision: the start of every task in each transcript, for the breakdown by difficulty, and the two independent coding passes behind the schema-strategy labels in Table~\ref{tab:qual_behavior}. Both annotations and their evidence, like the scripts, are in the replication package (see Data Availability).

\section{Results}\label{sec:results}

We report quantitative outcomes for all 20 participants first (Section~\ref{sec:overall} covers the mixed-effects models), followed by behavioral coding of the 11 sessions with usable think-aloud audio (Section~\ref{sec:qualitative}). Across the study, SQL-LLM cut per-query completion time by 31\% on the log scale ($p = 0.03$; about $212$ s on the raw scale, $p = 0.04$--$0.07$; Section~\ref{sec:overall}), but did not improve accuracy: SQL-LLM users answered $46\%$ of queries correctly versus $64\%$ for Snowflake ($p = 0.08$, 95\% CI $[-39, +2]$ percentage points). The think-aloud sessions show what changed behind those numbers: Snowflake users spent their effort exploring the schema and writing SQL from scratch, whereas SQL-LLM users offloaded initial schema lookup to the model, especially on easier tasks, and focused on checking the generated query.

\subsection{Quantitative Analysis}

We begin with per-query completion times and accuracy, then examine how errors occurred, how time and accuracy combine into overall throughput, and how the mixed-effects model tests these differences across tasks.

\begin{figure}[t]
    \centering
    \includegraphics[width=\linewidth]{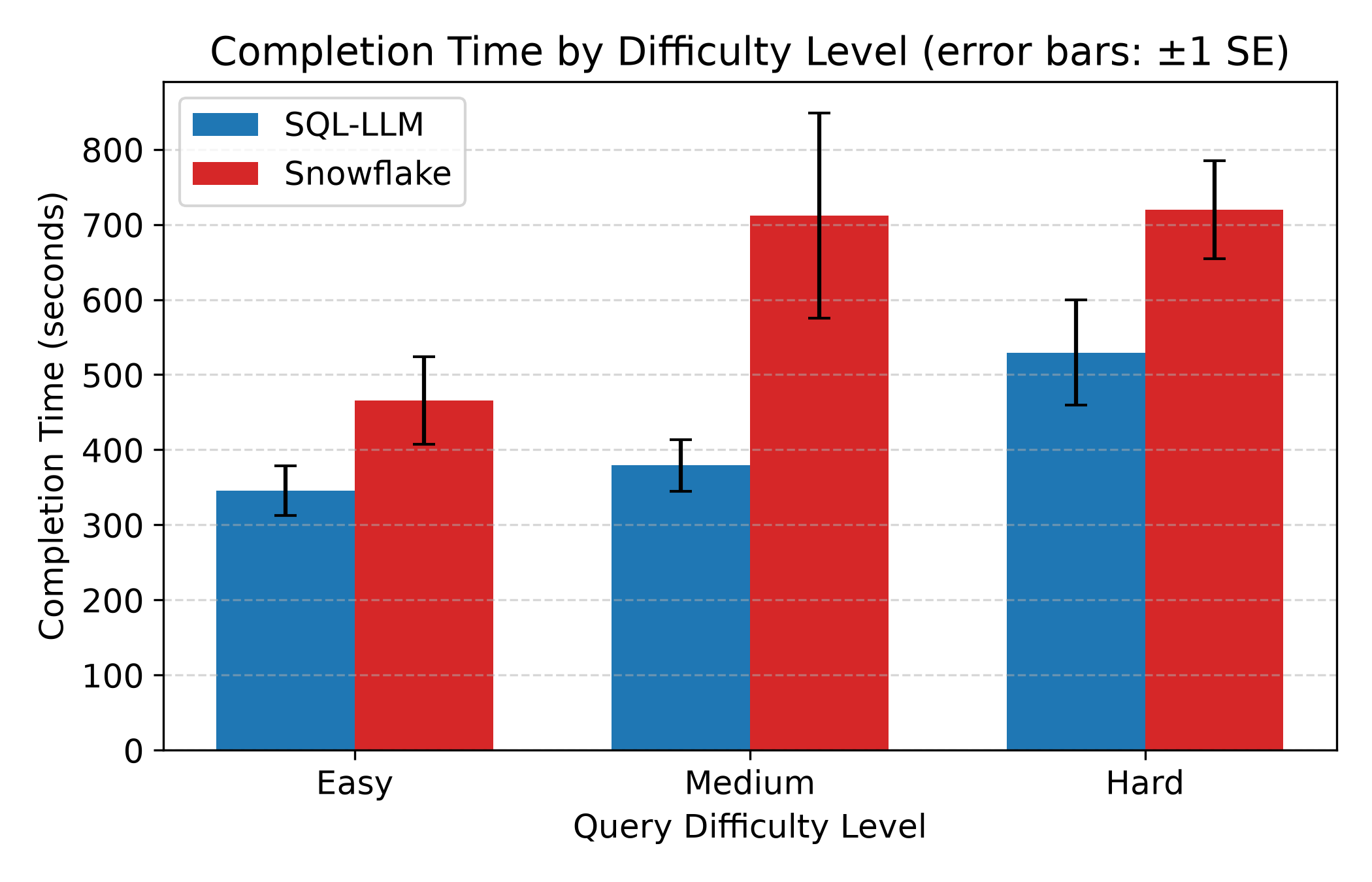}
    \caption{Mean query completion time (s) by difficulty level (Easy, Medium, Hard) for SQL-LLM and Snowflake. Error bars are $\pm 1$ standard error of the mean, computed over the task observations in each difficulty $\times$ group cell ($n = 40$ per cell, less any missing observations).}
    \label{fig:completion_time}

\end{figure}

Per-query completion times (Table~\ref{tab:results}, Appendix~\ref{app:suppfigs}) are lower for SQL-LLM on 11 of the 12 tasks, with the largest gaps on B-Q1 and B-Q3. Harder tasks take longer while SQL-LLM stays faster at every difficulty level (Figure~\ref{fig:completion_time}); whether that advantage widens reliably with difficulty is taken up by the linear mixed-effects model in Section~\ref{sec:overall}.

Faster completion did not improve accuracy. SQL-LLM users answered 46\% of queries correctly (55 of 120) versus 64\% for Snowflake (77 of 120) when executed against the gold BIRD database (Section~\ref{subsec:measures}). That 18-point difference favors Snowflake, though it falls just short of conventional significance at $N = 20$: a participant-level Welch test gives $p = 0.08$ (95\% CI $[-39, +2]$ percentage points), and a participant-level permutation test gives $p = 0.09$. A logistic regression with task fixed effects and participant-clustered standard errors reaches the same conclusion, estimating the odds of a correct answer under SQL-LLM at $0.44$ times those under Snowflake (95\% CI $[0.18, 1.08]$, $p = 0.07$). Because the confidence interval spans from rough parity down to a 39-point deficit, our data cannot rule out a substantial accuracy loss.

One session produced 11 of the SQL-LLM arm's 65 wrong answers. For one SQL-LLM participant the system returned ``[Timed Out]'' on 11 of 12 tasks, although the participant spent four to seven minutes on each; these non-answers are system failures, not wrong queries, and only one other non-answer occurred in either arm. We keep them in the primary analysis, because the treatment is the deployed SQL-LLM system, failures included. Without that participant the gap narrows to 50\% versus 64\% (odds ratio $0.53$, 95\% CI $[0.23, 1.24]$, $p = 0.14$; participant-level permutation $p = 0.15$), and treating the non-answers as missing gives the same odds ratio ($p = 0.13$). The direction holds in every variant; its size and the evidence for it shrink.

\begin{figure}[t]
    \centering
    \includegraphics[width=\linewidth]{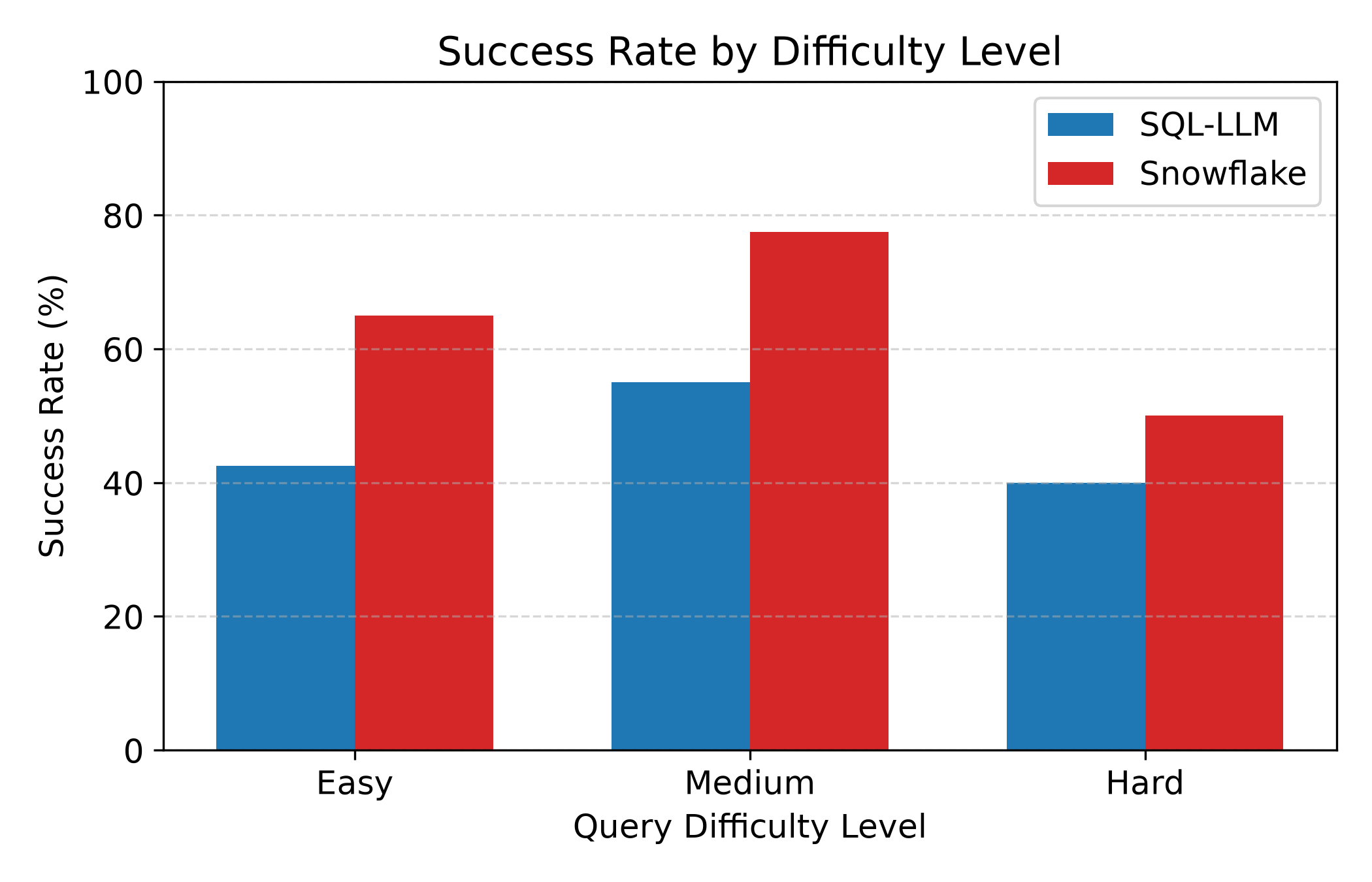}
    \caption{Queries answered correctly (\%) by difficulty, graded by executing each submission against the gold BIRD database and comparing to the gold answer (three independent LLM judges agree with the execution grade on 87.5\% of queries).}
    \label{fig:success_rate}
\end{figure}

Snowflake leads at every difficulty level (Figure~\ref{fig:success_rate}): Easy (65\% versus 42.5\%), Medium (77.5\% versus 55\%), and Hard (50\% versus 40\%), where both arms fall. We did not test the per-difficulty gaps for significance. The kind of failure matters as much as the count. Most wrong answers in both arms ran cleanly and returned a plausible, non-empty result that did not match the gold (34 of the 65 wrong SQL-LLM submissions, 26 of the 43 wrong Snowflake submissions). Nothing on the screen flags such an answer; only a check of the query against the question can catch it. Most of the rest were visible failures: execution errors (11 and 12), empty results (5 and 2), and non-answers (11 and 1). The remaining six (4 and 2) queried tables from a different BIRD ingest that our offline grader could not map; they may have run normally in the study warehouse, so we cannot say whether they failed visibly or quietly.

\paragraph{Auditing the gold answers.} Because execution grading presupposes the gold is right, and BIRD question--SQL pairs carry documented noise (15--49\% depending on domain, with 5--40\% of gold queries themselves wrong~\cite{wretblad2024understanding}), we audited all 12 gold queries against the prompts participants saw. The audit caught one outright mismatch (M-Q2, which had been paired with the wrong BIRD question) and one discrepancy between a task hint and its gold query (B-Q2, where the hint specified \texttt{author\_name LIKE 'Adam'}, excluding a hyphenated first name that the gold query's prefix match included; Appendix~\ref{app:goldaudit}). Our primary accuracy figures (45.8\% vs.\ 64.2\%) already use the corrected M-Q2 gold query, which was fixed before final grading, and neither audit issue affects the completion-time model in Table~\ref{tab:overall}. Grading B-Q2 under the literal hint instead of the BIRD gold query is a sensitivity check: it widens the accuracy gap to 46.7\% versus 68.3\% (exact participant-level permutation test, $p = 0.050$). Under either grading, SQL-LLM offers no accuracy gain. B-Q2 also illustrates how silent errors slip through: asked ``How many authors are named Adam?'', 4 of the 10 SQL-LLM users submitted a wildcard substring match (\texttt{LIKE} or \texttt{ILIKE '\%Adam\%'}), which also matches authors with the last name Adams (such as Abigail, Ansel, and Douglas Adams) and returns 39 instead of 21. The query runs without error and returns a believable count, giving no hint that it answered the wrong question.

\paragraph{Completion time by presentation position.} Completion times show no clear speedup across the 12-task sequence (only a slight downward slope for Snowflake; Appendix~\ref{app:suppfigs}). Because every participant received the 12 tasks in the same listed order, however, task position is confounded with task identity (Appendix~\ref{app:suppfigs}), so the session trace cannot separate practice from task difficulty. The think-aloud transcripts likewise show no shift in prompting strategy over the session (Section~\ref{sec:qualitative}).

\paragraph{Query reformulations.}
Reformulation showed no detectable per-minute difference across arms. Over the coded sessions, SQL-LLM users averaged $M = 6.5$ reformulations ($SD = 1.5$) and Snowflake users $M = 9.1$ ($SD = 3.1$); the totals are per-session, and normalized per audited minute the two arms are indistinguishable ($p = 0.75$). A reformulation follows the symmetric definition in Section~\ref{subsec:taskdesign}. The counts come from the double-coded transcript subset of Section~\ref{sec:qualitative} (two-coder Krippendorff's $\alpha = 0.66$).

\subsubsection{Overall Comparison}
\label{sec:overall}

Because every participant completed all 12 tasks, the data have a repeated-measures structure that a simple two-sample t-test ignores. The analysis therefore rests on a linear mixed-effects model (LME) with participant as a random intercept and Group, Difficulty, and Database as fixed effects, fit by restricted maximum likelihood (REML) to all 238 observations (two missing from the raw log). Table~\ref{tab:overall} reports the key fixed-effect estimates; Table~\ref{tab:comparison} summarizes results across all measured dimensions.

\begin{table}[t]
    \centering
    \begin{tabular}{lcc}
        \toprule
        \textbf{Metric} & \textbf{SQL-LLM} & \textbf{Snowflake} \\
        \midrule
        \multicolumn{3}{l}{\textit{Descriptive (per-participant means)}} \\
        Mean time (s) & 418.1 & 628.7 \\
        SD & 154.1 & 295.7 \\
        \midrule
        \multicolumn{3}{l}{\textit{LME fixed effect: Group (ref = Snowflake)}} \\
        $\hat{\beta}$ (s) & \multicolumn{2}{c}{$-212$} \\
        SE & \multicolumn{2}{c}{$106$} \\
        $z$ & \multicolumn{2}{c}{$-2.01$} \\
        $p$ & \multicolumn{2}{c}{$0.044$} \\
        \midrule
        \multicolumn{3}{l}{\textit{LME fixed effects: Difficulty (ref = Easy)}} \\
        Medium ($\hat{\beta}$, $p$) & \multicolumn{2}{c}{$+132$ s,\ $p = 0.059$} \\
        Hard ($\hat{\beta}$, $p$) & \multicolumn{2}{c}{$+216$ s,\ $p = 0.002$} \\
        \midrule
        \multicolumn{3}{l}{\textit{Database effects (ref = Books)}} \\
        Mondial Geo ($p$) & \multicolumn{2}{c}{$0.832$} \\
        Legislator ($p$) & \multicolumn{2}{c}{$0.728$} \\
        \midrule
        \multicolumn{3}{l}{\textit{Robustness of the Group effect (small-sample and log-scale)}} \\
        $t(18)$ reference for the Wald statistic & \multicolumn{2}{c}{$p = 0.059$, CI $[-434, +9]$ s} \\
        Welch / permutation, participant means & \multicolumn{2}{c}{$p = 0.066$ / $p = 0.045$} \\
        Cluster bootstrap CI, participant means & \multicolumn{2}{c}{$[-421, -32]$ s} \\
        Log-time LME: $-31\%$ per query & \multicolumn{2}{c}{$t(18)$ $p = 0.029$, CI $[-51\%, -4\%]$} \\
        Task fixed effects (raw / log time) & \multicolumn{2}{c}{$p = 0.064$ / $p = 0.031$} \\
        Excluding the extreme B-Q1 observation & \multicolumn{2}{c}{$\hat{\beta} = -174$ s, $p = 0.042$} \\
        \bottomrule
    \end{tabular}
    \captionsetup{skip=5pt}
    \caption{LME fixed-effect estimates for completion time. Model: Time $\sim$ Group + Difficulty + Database + (1$|$Participant); $n = 238$ observations (of the 240 possible from 20 participants $\times$ 12 tasks, two were missing from the raw log), fit by REML. Participant random-intercept SD $\approx 201$ s. The robustness block re-tests the Group contrast with participant-level degrees of freedom ($t$ with $18$ df, since treatment is assigned to 20 participants), distribution-appropriate refits, task fixed effects in place of the Difficulty and Database covariates, and outlier exclusion; the reanalysis script (\texttt{robustness\_reanalysis.py}) is in the replication archive.}
    \label{tab:overall}
\end{table}

Controlling for task difficulty and database, the raw-scale LME estimates that SQL-LLM users spent $212$ s less per query on average ($\hat{\beta} = -212$, SE $= 106$; Table~\ref{tab:overall}). Because interfaces were assigned to 20 participants rather than 238 independent tasks, the standard Wald $z$-test ($p = 0.044$) is slightly optimistic in a small sample; evaluated against participant-level degrees of freedom ($t(18)$) or via Welch, permutation, and cluster-bootstrap checks, the raw-time reduction sits right at the significance boundary ($p = 0.042$--$0.066$, Table~\ref{tab:overall}). Raw completion times are also heavily right-skewed, including one 5,456 s Snowflake outlier on B-Q1. We therefore treat the log-time model as primary: on the log scale, SQL-LLM reduces completion time per query by 31\% ($\hat{\beta} = -0.38$ log-seconds, $t(18)$ $p = 0.029$, 95\% CI $[-51\%, -4\%]$), and that reduction holds when we replace the difficulty and database covariates with task fixed effects or drop the B-Q1 outlier. Every specification points in the same direction, with a clear effect on the log scale and a borderline effect on the raw scale.

Hard tasks took significantly longer than Easy tasks ($\hat{\beta} = +216$ s, $p = 0.002$, 95\% CI $[+79, +353]$ s); database choice had no significant effect ($p > 0.7$). The large participant random-intercept SD ($\approx 201$ s) reflects substantial individual variation in baseline speed, which the mixed model partitions from the interface effect. The Group $\times$ Difficulty interaction was not statistically significant ($p > 0.1$), so the data do not support the gap widening with difficulty, though it is directionally larger for Medium queries.

\paragraph{Throughput.}
Because a user can finish faster simply by accepting a plausible but wrong query, looking at speed or accuracy alone can be misleading. We therefore combine the two into a single throughput measure: correct answers per hour of active on-task time. Across participants, throughput averaged 4.6 correct answers per hour for SQL-LLM and 4.2 for Snowflake (Welch $p = 0.77$; permutation $p = 0.78$; pooled across each arm, one correct answer took 912 s of on-task time under SQL-LLM versus 992 s under Snowflake). On net throughput, the two interfaces are nearly tied: SQL-LLM changed how participants spent their time, not how many correct answers they produced per hour. Within both arms, correct submissions were actually faster on average than incorrect ones (SQL-LLM 374 vs.\ 456 s; Snowflake 613 vs.\ 662 s), suggesting that wrong answers came mainly from tasks where participants got stuck rather than from hasty acceptance.

\begin{table}[t]
    \begin{center}
    \renewcommand{\arraystretch}{1.5}
     \resizebox{0.99\linewidth}{!}{
    \begin{tabular}{|l|c|c|p{4cm}|}
        \hline
        \textbf{Metric} & \textbf{SQL-LLM} & \textbf{Snowflake} & \textbf{Difference} \\
        \hline
        \multicolumn{4}{|l|}{\textit{Quantitative outcomes: all 20 participants ($n = 10$ per arm)}} \\
        \hline
        \textbf{Completion time} & 418.1 (154.1) s & 628.7 (295.7) s & $-31\%$ log scale ($p = 0.03$); $-212$ s raw ($p = 0.04$--$0.07$) \\
        \textbf{Accuracy} & 46\% & 64\% & $-18$ pts, $p = 0.08$, CI $[-39, +2]$ pts \\
        \textbf{Correct answers per hour} & 4.6 & 4.2 & Near parity ($p = 0.77$) \\
        \hline
        \multicolumn{4}{|l|}{\textit{Behavioral counts: 11 coded transcripts (4 SQL-LLM, 7 Snowflake), double-coded, $\alpha = 0.90$}} \\
        \hline
        \textbf{Query reformulations} & 6.5 (1.5) & 9.1 (3.1) & $p = 0.75$; $\alpha = 0.66$, below threshold \\
        \textbf{Frustration episodes} & 6.4 (10.3) & 4.5 (5.4) & $p = 0.82$; $\alpha = 0.99$, reliable \\
        \textbf{Schema checks} & 14.6 (4.8) & 25.7 (9.7) & $p = 0.32$; $\alpha = 0.78$, tentative \\
        \textbf{Hesitation episodes} & 3.8 (1.8) & 7.4 (3.2) & $p = 0.63$; $\alpha = 0.20$, below threshold \\
        \textbf{Expressions of confusion} & 10.0 (9.9) & 16.2 (7.2) & $p = 0.36$; $\alpha = 0.94$, reliable \\
        \textbf{Positive confirmations} & 9.5 (0.4) & 13.9 (1.4) & $p = 0.89$; $\alpha = 0.81$, reliable \\
        \textbf{Schema exploration (0--3)} & 1.8 (range 1--3) & 2.7 (range 2--3) & Snowflake explored more upfront; Table~\ref{tab:qual_behavior} \\
        \hline
    \end{tabular}
    }
    \end{center}
    \captionsetup{skip=5pt}
    \caption{Comparison of SQL-LLM vs.\ Snowflake across measured dimensions, with the sample behind each block stated in its divider row. Numeric cells are mean (SD). Behavioral cells are per-session event counts from the $n = 11$ transcripts, independently double-coded by two analysts (overall Krippendorff's $\alpha = 0.90$). Raw counts scale with session length; Section~\ref{sec:qualitative} describes the per-minute comparison. For each behavioral row, the Difference column reports the per-minute between-arm comparison and the category's two-coder Krippendorff's $\alpha$, with its conventional tier~\cite{krippendorff2004content} ($\alpha \geq 0.80$ reliable; $0.667 \leq \alpha < 0.80$ tentative; below $0.667$ too unreliable for conclusions). No category shows a detectable difference (all $p \geq 0.3$); the two categories below threshold carry no weight in our account. The last row is the mean schema-exploration level from the strategy coding of Table~\ref{tab:qual_behavior}.}
    \label{tab:comparison}
\end{table}

Table~\ref{tab:comparison} sets the dimensions side by side: faster completion without an accuracy advantage, near-parity throughput, and behavioral counts with no detectable per-minute difference in any category (normalization in Section~\ref{sec:qualitative}). We read this as the interfaces relocating the observable work: SQL-LLM users left much of the schema work to the model and checked its output, while Snowflake users explored the schema and built queries by hand.

\subsubsection{Per-Query Tests and Power}
Per-query independent two-sample $t$-tests, reported with effect sizes and confidence intervals in Appendix~\ref{app:perquery}, check whether the aggregate Group effect rests on one or two queries: point estimates favor SQL-LLM on 11 of 12 tasks, but at $n = 10$ per arm only B-Q3's interval excludes zero, so these tests are exploratory and carry no inferential weight of their own. The formal power analysis (also Appendix~\ref{app:perquery}) sizes the design at medium-to-large effects: a two-sample test on participant means has roughly 47\% power at the observed effect, which is why the corrected participant-level inference in Table~\ref{tab:overall} hovers near the significance boundary on the raw scale.

This power profile shapes how the null findings should be read. The behavioral \emph{rates} (events per audited session minute, compared in Table~\ref{tab:comparison}) showed no detectable between-arm differences in the $n = 11$ behavioral subset. We frame these as ``no detectable difference at this $N$'' rather than as evidence of equivalence between interfaces: distinguishing ``the interfaces shift the same total effort'' from ``the interfaces produce subtly different counts we cannot resolve here'' would take a much larger follow-up. At the observed effect sizes, detecting the largest per-minute differences (schema checks and confusion, Cohen's $d \approx 0.56$) with 80\% power needs about 52 participants per arm, and the smaller differences would need hundreds.

\subsection{Qualitative Analysis}
\label{sec:qualitative}

The aggregate outcomes told us that SQL-LLM users were faster, not what changed in their work. Think-aloud transcripts and screen recordings, available for 11 of 20 participants (4 SQL-LLM, 7 Snowflake), let us ask where the effort went. We read each session along three dimensions (confidence in query output, schema discovery strategy, and error recovery) and counted events in the six coded categories below. Under SQL-LLM, users left much of the schema navigation to the model and turned to checking the generated SQL; under Snowflake, users did their schema work earlier, exploring tables and building queries by hand. The coded record shows where the observable work went, not that total effort was equal on each side (per-minute rates below).

\paragraph{Coding methodology and inter-coder reliability.}
Two analysts independently coded all 11 transcripts across six categories: frustration episodes, schema-check events, hesitation episodes, reformulations, expressions of confusion, and positive confirmations. Both were hired for the task and blind to the study's hypotheses, to each other's coding, and to any prior counts. Transcripts were relabeled in random order, though the interaction modality (writing SQL vs.\ reviewing generated SQL) is evident from the transcript itself. Coding was conservative and event-based: each continuous stretch of one activity counted as a single event, ambiguous utterances were not counted, and each utterance was assigned its single best-fit category. Pooled across categories, two-coder Krippendorff's $\alpha$ was $0.90$ (interval metric, cluster-bootstrap 95\% CI $[0.81, 0.93]$; mean across the six categories $0.73$), with per-category $\alpha$ ranging from $0.99$ (frustration) and $0.94$ (confusion) down to $0.20$ (hesitation); schema checks, the category our account leans on, reached $\alpha = 0.78$. We report the two-coder mean counts and weigh each category by its reliability: frustration, confusion, and positive confirmations meet the conventional $\alpha \geq 0.80$~\cite{krippendorff2004content}; schema checks are tentative; hesitation ($\alpha = 0.20$) and reformulation ($\alpha = 0.66$) fall below the $0.667$ floor, so our account does not rest on them. Two transcripts (P2, P16) had sparse audio and yielded conservative lower bounds. Quotations are lightly edited to correct automatic-transcription artifacts, and participants appear under the neutral codes from Section~\ref{subsec:participants}. The full rubric, both coders' event logs, the reliability script, per-arm group means, per-minute rates, and representative events are in the replication archive (\texttt{human\_irr.py}, \texttt{human\_coding\_*.csv}, \texttt{coding-kit/CODEBOOK.md}, \texttt{coding\_examples.md}).

We also ran three independent LLM coders over the same transcripts; their annotations and their agreement with an earlier single-coder pass are reported in Appendix~\ref{app:llmcoders}.

Because raw event counts accumulate over a session, longer sessions naturally produce higher totals, and Snowflake sessions ran longer on average than SQL-LLM sessions ($129$ vs.\ $81$ audited minutes). Dividing each participant's counts by their audited session duration removes that confound. Once normalized to events per minute, none of the six coded behaviors differs significantly between arms (all Welch $p \geq 0.3$; the same holds among the seven sessions whose transcript timestamps needed no video correction). Positive confirmations illustrate why normalization matters: both human coders and all three LLM coders counted more total confirmations per session under Snowflake, yet the per-minute rates are nearly identical ($p = 0.89$). Snowflake users built queries clause by clause and murmured brief confirmations as each piece ran (``okay, looks good'', ``there we go''), whereas SQL-LLM users received a complete query at once and confirmed at the end.

To check whether the 4-versus-7 sample split (Section~\ref{subsec:procedure}) masks a difference, we re-ran the per-minute comparison on all $\binom{7}{4} = 35$ balanced subsamples of four SQL-LLM versus four Snowflake sessions. No category reached significance under an exact permutation test in any of the 35 subsamples (an unadjusted $t$-test crossed $p < 0.05$ in 3 of 210 tests, all for schema checks). The null result also holds when evaluating each coder separately and when dropping any single session, except that removing the longest Snowflake session (273 minutes) brings schema checks to $p = 0.046$. Two categories nonetheless show a consistent direction: per-minute schema checks were higher under Snowflake in 34 of the 35 balanced subsamples, and confusion was higher in 33, whereas the other four categories flipped sign across subsamples.

\subsubsection{Qualitative Metrics}

The counts live in Table~\ref{tab:comparison}; we read them as a map of where each arm's observable activity fell. This subsection leads with schema interaction, the category on which the account turns.

The finding our account rests on is \textit{when} schema interaction occurred, more than how often. Snowflake users explored the schema before writing a query: six of the seven browsed tables in the editor's object panel, and one (P12) queried \texttt{INFORMATION\_SCHEMA}. SQL-LLM proposed the tables relevant to each question and let the user pick from them, and two of its four coded users (P1, P9) then checked specific column names mainly \textit{after} receiving the generated SQL; P2 mixed early looks with later checks, and P7 opened and read tables before most generations. Schema checks ran higher per minute under Snowflake in almost every balanced subsample but not significantly (Table~\ref{tab:comparison} and the balanced checks above), consistent with the model taking over the initial schema work while the analyst's own checking shifts downstream. The schema work did not disappear; it moved from construction toward verification.

Did SQL-LLM users check the schema more on harder tasks, where model errors are more likely? Because participants typically read each task number and prompt aloud (Section~\ref{subsec:measures}), we could assign 89\% of coded events to a specific task (nine transcripts mapped cleanly across all tasks, while two SQL-LLM transcripts, P2 and P7, spanned split recordings that could only be mapped for a subset of tasks: four SQL-LLM sessions cover Easy tasks and three cover Medium and Hard tasks). Treated descriptively, coded behaviors did not concentrate on Hard tasks. Instead, Snowflake's higher rate of schema checks came entirely from Easy and Medium tasks (0.24 vs.\ 0.21 and 0.31 vs.\ 0.24 events per minute) and disappeared on Hard tasks (0.18 vs.\ 0.19). Only hesitation ($\alpha = 0.20$) rose with task difficulty, increasing for five of the seven Snowflake users. In short, the model saved users from browsing the schema on easier tasks, but on Hard tasks SQL-LLM users checked the schema just as often as manual SQL users.

Did the visible verification pay off? Linking each coded task to its execution grade shows that voiced satisfaction during a task did not reliably mean the participant had arrived at the right answer. Our positive-confirmation code captures any expression of satisfaction during a task, whether confirming an intermediate step or accepting a final result. In the SQL-LLM arm, tasks with at least one positive remark were correct 57\% of the time (10 of 23 were wrong), compared with 38\% for tasks without one ($p = 0.43$); in the Snowflake arm, tasks with a positive remark were correct 65\% of the time (26 of 75 were wrong), compared with 67\% without one ($p = 1.0$). Schema checks occurred on nearly every task in both arms and therefore did not distinguish right answers from wrong ones either. Expressions of confusion, by contrast, marked trouble rather than success: in these descriptive task-level comparisons across the 11 participants, tasks with voiced confusion were solved less often in both arms (31\% vs.\ 67\% for SQL-LLM; 57\% vs.\ 88\% for Snowflake; separating final-answer confirmations from step-level ones is left to Section~\ref{sec:futurework}). In neither interface could a listener tell from a participant's positive remarks whether the final query was actually correct.

Frustration showed no detectable per-minute difference (Table~\ref{tab:comparison}); only its \textit{source} differed, arising under SQL-LLM from semantic mismatches between the user's intent and the model's generated query and from interface friction such as lag and difficulty interrupting generation, and under Snowflake from schema navigation failures and SQL syntax errors.

\subsubsection{Coded Behavioral Patterns}

The transcripts show what the shift felt like. On confidence, SQL-LLM users displayed early trust in the model's output, treating it as a first pass to check rather than rewrite, while Snowflake users worked by trial and error:

\begin{quote}
\textit{``Okay yes, the query given by the AI is correct. [...] I will just verify it quickly.''}
-- P1, SQL-LLM
\end{quote}

On schema discovery, Snowflake users explored before querying, mostly by browsing tables and trialing joins (one, P12, by querying the information schema), whereas SQL-LLM users explored less (Table~\ref{tab:qual_behavior}) and two of the four checked joins and columns mainly after the fact:

\begin{quote}
\textit{``What I'm going to do is take a look at these tables. [...] We can say select table name from information schema.''}
-- P12, Snowflake
\end{quote}

On error recovery, the same split held: Snowflake users showed frustration navigating ambiguous joins or failing queries, while SQL-LLM users debugged at the level of logical alignment rather than syntax.

Across the three themes the difference was strategic, not affective: both groups sustained attention and adjusted iteratively, and the interface set \textit{where} the work happened, not whether it happened. Table~\ref{tab:qual_behavior} records the per-participant coding behind this pattern.

\begin{table}[t]
    \centering
    \begin{tabular}{llll}
        \toprule
        \textbf{Participant} & \textbf{System} & \textbf{Exploration level} & \textbf{Primary method} \\
        \midrule
        P1  & SQL-LLM   & 1 Targeted check & Table browser \\
        P2  & SQL-LLM   & 2 Moderate       & Table browser \\
        P7  & SQL-LLM   & 3 Systematic     & Table browser \\
        P9  & SQL-LLM   & 1 Targeted check & Table browser \\
        P11 & Snowflake & 3 Systematic     & Table browser \\
        P12 & Snowflake & 3 Systematic     & Metadata queries \\
        P13 & Snowflake & 3 Systematic     & Table browser \\
        P14 & Snowflake & 2 Moderate       & Table browser \\
        P15 & Snowflake & 3 Systematic     & Table browser \\
        P16 & Snowflake & 2 Moderate       & Table browser \\
        P19 & Snowflake & 3 Systematic     & Table browser \\
        \bottomrule
    \end{tabular}

    \captionsetup{skip=5pt}
    \caption{Schema-discovery strategy of each coded participant, classified with one scheme for both arms: an ordinal level of schema exploration (0 Minimal, 1 Targeted check after a query exists, 2 Moderate exploration on some tasks, 3 Systematic exploration before querying on most tasks) and the primary method. Two independent coding passes labeled every session from the full transcript with quoted evidence; they agreed exactly on the level for 9 of 11 sessions (quadratic-weighted $\kappa = 0.83$, never more than one level apart) and on the method for all 11, and the two disagreements were settled from the evidence. Under the common scheme, each arm shows three distinct labels. Choosing tables in SQL-LLM's table picker did not count as exploration; opening tables or reading columns beyond the pick did. Affective dimensions are not tabulated per participant: their counts show no detectable between-arm difference (Table~\ref{tab:comparison}) and we have no validated rubric for per-person labels.}
    \label{tab:qual_behavior}
\end{table}

The transcripts add the spread the group means hide: frustration was mixed within the SQL-LLM group, concentrated in participants working complex multi-join queries or hitting interface lag, so the low-friction pattern of reviewing generated SQL held for most SQL-LLM users but not all.

We state the design implication as a hypothesis: SQL-LLM appears to reduce the upfront burden of schema navigation, mainly on easier tasks (shifting it downstream, since users still checked generated joins and column choices), and to add the burden of checking and correcting generated queries. Schema visibility, a readable generated query, and clear error recovery are the levers an organization can actually pull; interactive subquery previews, highlighted faulty clauses, or partial execution could align better with how users repair queries and reduce the effort of correction. Section~\ref{subsec:readiness} returns to query visibility and schema governance.

\subsubsection{Micro-level Repair Strategies}

The shift also showed up at the level of single clauses, in how users repaired a query once it went wrong. Snowflake users attempted localized adjustments, often with difficulty: P19 caught a missing grouping only after running the query, and P12 returned to \texttt{INFORMATION\_SCHEMA} repeatedly between attempts, re-checking table and column names with little system-level support for contextual feedback. In the episodes we observed, SQL-LLM users instead checked and adjusted at the semantic level, treating a system error as a prompt for incremental refinement rather than a reason to start over, isolating and checking one component at a time, as when P1 commented out part of a generated query to test the inner query on its own. Snowflake users were likelier to locate the fault in their foundations and rebuild rather than patch:

\begin{quote}
\textit{``Maybe I was just using the wrong table and I don't understand how the historical table works.''} -- P11, Snowflake
\end{quote}

The contrast is between trusting and refining a generated query and rebuilding one from scratch. In these episodes, SQL-LLM supported low-friction, micro-level revision, whereas the manual editor demanded more complete syntactic correctness upfront. These repair episodes are selected from 11 transcripts, not a census of repair behavior. Some strategies appear only in the SQL-LLM transcripts.

\subsubsection{Emergent Strategies in the SQL-LLM Arm}

Only four SQL-LLM participants produced codable think-aloud audio, so everything here is an observation from a small set, reported as raw counts rather than rates. Three strategies surfaced, unevenly: prompt-structure reuse, where a participant adapted a structure that had worked on an earlier task rather than phrasing each from scratch (one participant); deliberate probing of the system's flexibility by varying prompt specificity (one participant); and the refine-not-rebuild move from the previous subsection (two participants). We did not see the learning curve one might expect across successive tasks: the participant who reused prompt structures adopted the routine on an early task and kept it, so we read the variation as differences in habitual style that participants brought with them rather than a trajectory the session produced. What the interface did was let some users exercise a strategy they already had, which argues for aligning error feedback and interaction flow with the reuse, probing, and refinement we observed rather than an assumed learning curve. Whether users without SQL training would develop or benefit from the same strategies remains open (Section~\ref{sec:futurework}).

\section{Discussion}\label{sec:discussion}

In our study, SQL-LLM cut completion time per query by 31\% on the log scale relative to the manual-SQL arm (about $212$ s on the raw scale), a large speed effect. The nearest comparable study measured something else: Ning et al.~\cite{ning2023empirical} started every participant from model-generated SQL to repair and found that repair interfaces bought no significant overall gain in speed or accuracy over editing it directly. Read together, the two studies fit a simple account: the speed gain comes from generation itself, which Ning et al. held constant, not from the tools for repairing what the model wrote. The studies differ in tasks, systems, and participants, so the pairing suggests this account rather than tests it. Accuracy moved in neither study: our SQL-LLM users were no more correct than their Snowflake counterparts, and our point estimate runs 18 points lower, a possible loss that $N = 20$ can neither confirm nor exclude. On this account, speed comes from the model, while verification remains the user's problem, and our data give no sign that it has moved. The same reading may explain why our accuracy gap is larger than the non-significant one in the older paper-and-pencil study of Vassiliou et al.~\cite{Vassiliou1983}: no system ran there, so no plausible but wrong output reached a user to be checked (Section~\ref{sec:related}). The implications below turn on verification as the user's problem.

\subsection{What the Results Mean}

Our central reading is that SQL-LLM reallocated the work of querying rather than removing it: in the coded sessions, observable work moved from query construction toward output verification. Behavioral coding showed no detectable per-minute difference in any coded category (Section~\ref{sec:qualitative}); the faster completion time brought no gain in accuracy, and correct answers per hour stayed near parity. The same cost appears in AI-assisted programming, where verifying suggestions occupies a measurable share of programmers' time~\cite{mozannar2024reading}. The interface made querying faster, not more correct. For users who already have working SQL knowledge, it redistributes the analyst's time: faster turnaround per query, with a possible accuracy cost the present sample cannot pin down. This redistribution reframes what counts as a benefit for organizations evaluating NLIDB adoption: an argument built around ``less work''~\cite{davis1989tam} misreads the tool, and an adoption case must weigh the speed gain against an accuracy risk our data flag but do not settle.

The 46\% accuracy needs context, because on its face it reads as an indictment. Trained users making errors is the expected condition, not an anomaly: the literature documents high rates of syntactically valid but semantically wrong queries among students learning SQL~\cite{brass2006semantic, taipalus2018errors}, and even BIRD's own human baseline among database engineers and database students is 92.96\% rather than 100\%~\cite{li2024can}. That baseline covers different tasks, tools, and conditions and does not explain rates as low as 46\% and 64\%; we cite it only to establish that wrong-but-executable queries are the working condition of SQL, whoever writes it. What matters for adoption is who catches them, and our data leave open whether anyone does: would a user catch a query that is plausible but quietly wrong~\cite{parasuraman1997humans}, such as the B-Q2 substring query in Section~\ref{sec:results}? Our task-level data give a partial answer, and it is not reassuring: in both arms, tasks on which a participant voiced satisfaction were not detectably more often correct than tasks on which they did not (Section~\ref{sec:qualitative}). Until confidence in generated SQL is shown to track correctness, the prudent default is to keep the generated query where a reader can check it, which Section~\ref{subsec:readiness} states as a design hypothesis.

\paragraph{Verification as trust calibration.} The verification burden is a question of trust calibration in the sense of the automation literature. Lee and See~\cite{lee2004trust} call trust calibrated when it matches what the automation can actually do, and Parasuraman and Riley~\cite{parasuraman1997humans} name the two ways it fails: misuse (relying on automation that is wrong) and disuse (rejecting automation that is right). Our six coded behaviors map onto these terms. A positive confirmation is an act of reliance; whether reliance is calibrated is whether confirmations land on correct answers. Schema checks are the monitoring that calibrated reliance requires. A reformulation is a revision of trust after the user detects a failure. Confusion marks a gap in the user's model of the system or the data, the understanding on which Lee and See ground appropriate trust. Frustration is the affective response to a failure, and hesitation is the delay before committing. Read this way, our result points to poor calibration in both arms: positive remarks did not reliably separate right answers from wrong ones, so reliance was poorly calibrated whether the SQL came from a model or from the user's own hands. The caveat is that our code captures satisfaction voiced anywhere in a task, not a final verdict on the answer. What the NLIDB changes is the object of trust, a generated query in place of one's own; the transcripts suggest it does not change how well that trust is calibrated.

\paragraph{A simpler account.} A simpler reading fits our data as well: SQL-LLM users wrote less accurate queries and gained nothing from their checking. That reading is consistent with the accuracy interval, and the calibration result gives it no resistance. Two observations narrow it without refuting it. The failure of positive remarks to track correctness appears under Snowflake too, so weak verification is not specific to the natural-language interface. And part of the accuracy gap traces to one session in which the system timed out, a failure no verification could repair. The two accounts agree that work moved; they differ on whether the moved work protects correctness. Separating them needs a measure our design lacks: how often each interface's users catch a plausible but wrong query. Seeded wrong queries with per-interface catch rates, and logs of every step from generated to submitted SQL, would provide it (Section~\ref{sec:futurework}). If SQL-LLM users catch seeded errors as often as manual-SQL users, the moved effort still does its job; if they miss more, the simpler account is right about the mechanism of the accuracy loss.

\subsection{Implications for Adoption}
\label{subsec:readiness}

The implications below follow from our reading of the evidence, not from a test of deployment, and each inherits the limits stated in Section~\ref{sec:intro} and Section~\ref{subsec:limitations}. If the reallocation account holds, an organization that hands its analysts an NLIDB shifts a skill requirement rather than eliminating it: someone must read the generated SQL and confirm it means what the question asked. In the terms of technology-acceptance research, that dependency falls under facilitating conditions, which include having the knowledge needed to use the system, and on our reading it would govern who benefits, what has to change, and where deployment could start~\cite{venkatesh2003utaut}.

The analyst's job shifts from writing queries to reviewing them. In the think-aloud transcripts, SQL-LLM participants read the generated SQL against their intent: the work of writing queries gave way to the work of checking them. Review skill is necessary, though on our data not sufficient: participants checked the SQL on nearly every task and still missed many wrong answers. Adopting an NLIDB is therefore an investment in SQL review skill, not a way to retire SQL skill: staff still need to read a join, spot a wrong grouping, and recognize a result that looks plausible but is wrong. Training that pitches the tool as a way around SQL literacy removes the exact skill the workflow depends on.

Schema governance is the foundation that both the model and the human reviewer rely on. In our study, the model handled initial schema discovery mainly on the easier tasks, helped by small, clean schemas with descriptive table and column names. In an enterprise warehouse with hundreds of cryptic tables, both steps get harder: the model has more ways to pick the wrong join or column, and the reviewer has more work to check. On the model's half, there is direct evidence that schema-identifier naturalness correlates with LLM Text-to-SQL performance~\cite{luoma2025snails}. Well-documented, consistently named schemas keep generated SQL readable, whereas poor schema hygiene inflates the verification burden before any time savings can materialize.

Keeping the generated SQL visible and editable is a sensible design default. Because participants' positive remarks did not reliably distinguish right answers from wrong ones (Section~\ref{sec:qualitative}), hiding the SQL behind a pure chat interface would remove the primary check our SQL-literate participants used. Whether visible SQL measurably improves error detection remains a design hypothesis to test directly, for example by comparing catch rates on seeded errors when the SQL is shown versus hidden.

Rollout can be staged by who can verify the output. Analyst and data-engineering teams can capture the speed gain on routine queries because they can read and audit the generated SQL. Extending the same tool to staff who cannot read SQL is a different bet, because they lack the verification channel our participants relied on. Starting deployment with SQL-literate teams, as an accelerator rather than a replacement, matches the tool to the users who can verify its output.

\subsection{Limitations}
\label{subsec:limitations}

The interpretation is bounded twice: internally, by what a 20-participant design can statistically confirm, and externally, by the conditions we studied. Those conditions are also the axes along which the findings are most likely to vary elsewhere: team SQL literacy, schema design, task framing, the system we deployed, and the single-session snapshot we measured.

\begin{itemize}
    \item \textbf{Statistical power and sample size}: The between-subjects design splits 20 participants 10/10 and is powered to detect only medium-to-large effects. The completion-time result clears this bar on the log scale ($p = 0.03$) and hovers at the boundary on the raw scale ($p = 0.04$--$0.07$ across specifications; Table~\ref{tab:overall}). Several headline comparisons do not. The accuracy gap (46\% vs.\ 64\%, $p = 0.08$, CI $[-39, +2]$ points) shows no detectable gain for SQL-LLM, and an interval this wide cannot rule out a substantial loss, so the direction favoring Snowflake is plausible but not established. Part of the gap comes from one session in which the system timed out; without it the gap is 14 points ($p = 0.14$). The coded behaviors show no between-arm difference in per-minute rate; schema checks show a consistent direction across balanced subsamples (Section~\ref{sec:qualitative}), but not a significant one. Where we report no difference, read it as no detectable difference at this $N$. Establishing that accuracy is \emph{preserved} would require a noninferiority design with a prespecified margin and a larger sample.
    \item \textbf{Participant background and diversity}: All participants had at least basic SQL literacy; none were complete novices, and all came from data analysis backgrounds. This ensured fair comparisons between systems, but for users with little or no SQL experience, NLIDBs may yield greater benefits or different failure modes. The benefit for non-technical users that earlier work reports (Section~\ref{sec:related}) lies outside what our data can test. The SQL-education literature sharpens the concern: novices produce syntactically valid but wrong-result queries at high rates, with the same logical errors recurring across students~\cite{taipalus2018errors}, so the verification channel our SQL-literate participants relied on is exactly the channel a novice population would lack.
    \item \textbf{Task hints removed natural ambiguity}: The textual hints that disambiguated each task (Section~\ref{subsec:taskdesign}) reduced the natural ambiguity real users would encounter, and their net effect on the between-arm comparison is not identified. Results should be read as usability under \textit{disambiguated} task framing; performance under hint-free, naturalistic phrasing remains an open question.
    \item \textbf{Flat compensation}: Participants received a fixed \$100 with no accuracy-contingent component, so submitting a plausible but unverified answer carried no cost. This satisficing risk applies to both arms but is not symmetric: accepting a generated query and moving on is easier than abandoning a half-written one, so the accuracy comparison may partly reflect incentives rather than interfaces, and generalization to settings where analysts bear the cost of wrong answers is limited.
    \item \textbf{Behavioral coding reliability}: All 11 transcripts were independently double-coded by two analysts (pooled Krippendorff's $\alpha = 0.90$; mean across categories $0.73$), with agreement high for frustration, confusion, and positive confirmations ($\alpha = 0.81$--$0.99$), tentative for schema checks ($\alpha = 0.78$), at the $0.667$ boundary for reformulation ($\alpha = 0.66$), and low for hesitation ($\alpha = 0.20$). The coded subset is also a post-assignment selection (11 of 20, unevenly 4/7 across arms), and per-session event counts scale with session length, so we compare per-minute rates rather than raw totals. These limits bound how much weight the behavioral counts can bear.
    \item \textbf{Transfer to large, complex, real-world schemas}: Tasks covered Easy to Hard difficulty levels but ran on three small- to medium-scale public benchmark schemas (Books, Mondial Geo, Legislator). Enterprise databases with hundreds of tables, complex joins, and security constraints may present challenges these tasks did not surface.
    \item \textbf{A paradigm comparison, not a system benchmark}: We compared one deployed NLIDB against one conventional SQL editor. The findings speak to the NLIDB-versus-manual-SQL paradigm for SQL-literate users, not to how SQL-LLM ranks against other NLIDBs. A different backing model, prompting strategy, or interface could change the size of the speed effect and the shape of the verification burden, so we offer the reallocation account as a hypothesis about interfaces that expose an editable generated query, not a claim about this particular system.
    \item \textbf{Longitudinal effects}: The study measured immediate usability, a snapshot that cannot see how learning and workflow fit evolve with use, or how trust shifts as users gain experience with the system~\cite{hoff2015trust}.
\end{itemize}

\section{Future Work}\label{sec:futurework}

The study leaves one question above the rest: when the model produces a quietly wrong query, do users catch it? Section~\ref{sec:discussion} explains why our data cannot answer it. Most directions below attack that gap; one removes the framing aid our task hints supplied, and one re-runs the comparison on tasks no public corpus contains.

\begin{itemize}
    \item \textbf{Verification accuracy under seeded errors}: Inject syntactically valid but wrong-result queries, and measure how often each interface's users catch the error before accepting it. This turns a calibration problem our data can only describe into a measurable catch rate and false-confidence rate.
    \item \textbf{A finer rubric for confirmation behavior}: Our coding (Section~\ref{sec:qualitative}) does not separate final-result confirmations from the running step-confirmations that accumulate when a query is built by hand, so raw counts stay confounded with construction style. A rubric distinguishing the two would let confirmation behavior be compared on equal footing rather than through a count that mixes the two.
    \item \textbf{Verification without a readable query}: Our participants could read and check a generated query before accepting it. Study a population that cannot read it: with the verification channel closed, does confidence run ahead of correctness, and does anything replace the check? This tests whether the reallocation account holds when its verifying step is unavailable.
    \item \textbf{Hint-free, naturalistic phrasing}: Re-run the task set without the textual hints that disambiguated phrases such as ``most orders,'' the framing aid Section~\ref{subsec:limitations} flags as a limitation. Removing them restores the intent ambiguity real users face and isolates how much of the construction saving depends on disambiguated framing.
    \item \textbf{Privately authored tasks}: BIRD and Spider are public corpora the backing model may have seen in training (Section~\ref{sec:sqlllm-ui}). A confirmatory study should use privately authored tasks over schemas absent from public corpora, logging the model snapshot, prompts, and validator behavior. This would settle whether the headline comparison survives on tasks the model cannot have seen.
    \item \textbf{Instrumented interaction logging}: Capture the initial generated SQL, every edit, regeneration, chat turn, and validator rejection. These traces would measure the path from generated to submitted query directly (initial model accuracy, how often users repair a wrong query or break a right one), turning a mechanism this paper infers from transcripts into a logged, quantifiable process.
\end{itemize}

\section{Conclusion}\label{sec:conclusion}

A natural-language database interface does not remove the work of querying; it shifts the visible work from writing SQL to checking it. Within the limits of our study (20 SQL-literate participants, 11 with coded think-aloud audio, and task hints that named target columns and operators), SQL-LLM cut completion time per query by 31\% compared with writing SQL by hand in Snowflake (Section~\ref{sec:overall}). Yet that speed gain brought no accuracy advantage: SQL-LLM users answered 46\% of queries correctly versus 64\% for Snowflake, an 18-point gap this sample cannot confirm ($p = 0.08$, or 14 points without one session where the system timed out), leaving correct answers per hour nearly identical across interfaces. In the think-aloud sessions, the model absorbed initial schema lookup, mainly on the easier tasks, while users shifted their attention to checking the generated SQL. Our evidence therefore supports a reallocation of querying effort and time per query, not a reduction in work, and it does not show that the moved work protects correctness.

For organizations considering an NLIDB, the takeaway depends on who uses the tool. Our participants knew SQL and could inspect the queries the model wrote; users without SQL training cannot fall back on that check.

That channel pays off when trust in it is calibrated to how reliable the generated SQL actually is~\cite{lee2004trust}, so that analysts believe correct output and doubt incorrect output. In our coded tasks, participants' expressions of satisfaction were a poor guide to whether their final answer was right under either interface. For SQL-literate teams, the central question for deployment is not how fast the model writes a query, but whether users catch the queries that are quietly wrong.

\section*{Ethics Statement}
This study involved human participants recruited via Upwork.
All participants provided informed consent,
and their personal information was anonymized.
The study protocol was reviewed and approved by the
Institutional Review Board of New York University
(IRB-FY2023-7595).

\section*{Conflicts of Interest}
SQL-LLM is a deployed system that we evaluate under a
pseudonym, a condition of the anonymity promised to its
provider. The authors have no financial interest in, and no
professional affiliation with, the provider of SQL-LLM or
with Snowflake.
Snowflake was chosen as a representative SQL platform,
which, at the time of the experiment, did not include any
Text-to-SQL capabilities in its Web UI; the comparison evaluates
interface paradigms, not vendors.

\section*{Data Availability}
The analysis code and study data (think-aloud transcripts,
submitted queries, timing logs, behavioral codings, the
execution-grading pipeline, and the robustness-reanalysis
script behind Table~\ref{tab:overall}) are available as a
replication package on Zenodo:
\url{https://doi.org/10.5281/zenodo.21433589}.
Participants appear throughout the package under the same
neutral codes (P1--P20) used in this paper; recordings and
raw files that could identify individuals are excluded.
Where an artifact does not exist, we say so rather than
omit silently: the exact GPT-4o snapshot identifier is not
recoverable, and validator rejections, fine-grained
interaction traces, and the open-ended debrief responses
were not captured or are not analyzable (Sections~\ref{sec:sqlllm-ui}
and~\ref{subsec:procedure}). Appendix~\ref{app:preprint} lists the
material changes from the first preprint version of this work.

\appendix

\section{Supplementary Quantitative Material}
\label{app:suppfigs}

Table~\ref{tab:results} reports per-query completion times; the aggregate model is in Section~\ref{sec:overall}.

\begin{table}[t!]
    \centering
    \begin{tabular}{llcc}
        \toprule
        \textbf{Query} & \textbf{Difficulty} & \textbf{SQL-LLM (s)} & \textbf{Snowflake (s)} \\
        \midrule
        B-Q1 & Medium & 402.9 $\pm$ 255.2 & 1103.9 $\pm$ 1553.9 \\
        B-Q2 & Easy   & 275.6 $\pm$ 184.3 & 269.8 $\pm$ 62.6 \\
        B-Q3 & Medium & 474.7 $\pm$ 268.0 & 803.9 $\pm$ 375.5 \\
        B-Q4 & Hard   & 364.8 $\pm$ 130.1 & 634.7 $\pm$ 385.9 \\
        M-Q1 & Medium & 395.9 $\pm$ 139.8 & 615.1 $\pm$ 341.6 \\
        M-Q2 & Hard   & 291.5 $\pm$ 107.8 & 410.3 $\pm$ 247.8 \\
        M-Q3 & Hard   & 893.1 $\pm$ 657.2 & 978.6 $\pm$ 385.6 \\
        M-Q4 & Easy   & 291.6 $\pm$ 290.3 & 484.3 $\pm$ 365.6 \\
        L-Q1 & Easy   & 386.4 $\pm$ 136.2 & 442.1 $\pm$ 168.9 \\
        L-Q2 & Medium & 242.8 $\pm$ 116.8 & 313.7 $\pm$ 218.7 \\
        L-Q3 & Hard   & 568.3 $\pm$ 391.5 & 824.8 $\pm$ 398.9 \\
        L-Q4 & Easy   & 429.8 $\pm$ 183.2 & 664.9 $\pm$ 577.5 \\
        \bottomrule
    \end{tabular}

    \captionsetup{skip=5pt}
    \caption{Per-query completion time (mean $\pm$ SD, in seconds) for SQL-LLM and Snowflake, with each query's difficulty level.}
    \label{tab:results}
\end{table}

\begin{figure}[t]
    \centering
    \includegraphics[width=\linewidth]{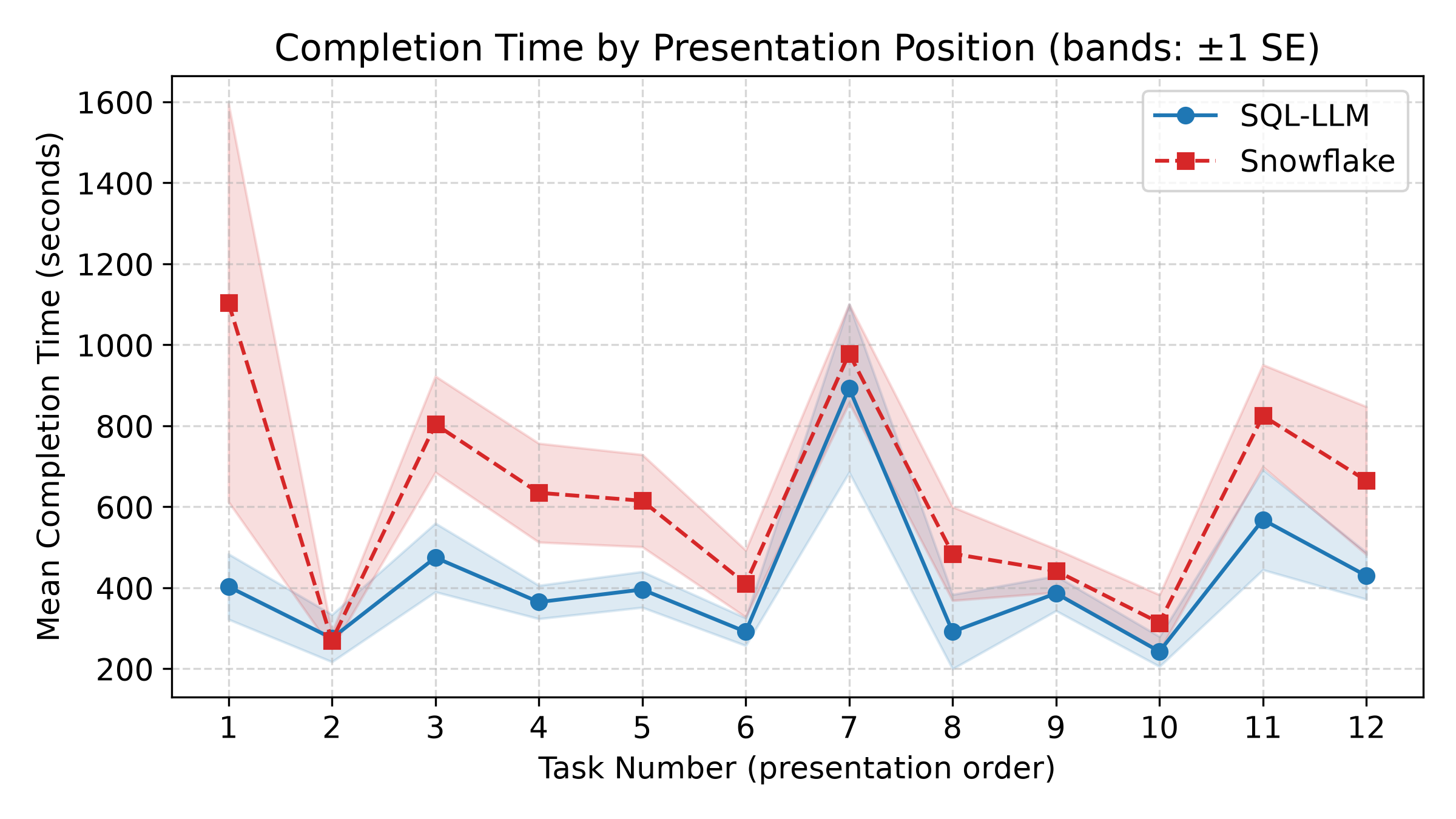}
    \caption{Mean completion time (s) by presentation position across the 12-task sequence, by group, with $\pm 1$ standard-error bands. Position is confounded with task identity (all participants saw the same tasks in the same listed order), so these traces describe the session profile and are not a learning curve.}
    \label{fig:learning_curve}
\end{figure}

\paragraph{Completion time by presentation position.}
The data neither establish nor rule out a within-session practice effect, because task position is confounded with task identity: every participant saw the same 12 tasks in the same listed order, so the mean time at each position reflects that position's specific task (its difficulty and database) as much as any practice. What we can report is the descriptive pattern by \emph{presentation position} (Figure~\ref{fig:learning_curve}). A linear fit of mean completion time against position is essentially flat for SQL-LLM (about $+6$ s per task, $r = 0.12$) and weakly negative for Snowflake (about $-14$ s per task, $r = -0.20$), with neither slope distinguishable from zero and no positional trend in accuracy. Measuring learning would require randomized or counterbalanced task order.

The apparent early drop in the Snowflake curve is driven largely by a slow, high-variance first task rather than by steady speed-up, and the SQL-LLM curve does not compress over the 12 positions; given the position--task confound, the flat traces cannot rule out learning that the task mix masks. The qualitative record (Section~\ref{sec:qualitative}) is consistent: we saw no shift from instance-specific to more abstract prompting as a session progressed.

\section{Gold-Answer Audit}
\label{app:goldaudit}

Grading against BIRD gold answers presupposes the gold is right, and an audit of BIRD found noise in 15--49\% of question--SQL pairs depending on domain, with 5--40\% of gold queries themselves wrong~\cite{wretblad2024understanding}. We therefore audited all 12 gold queries against the task prompts participants actually saw, cross-checked against the think-aloud recordings, and caught one outright mismatch. Our grading pipeline had matched M-Q2 to the BIRD question ``Which country has the lowest inflation rate?'', but participants were shown a different BIRD question over the same tables (``What is the name of the country whose citizens have the lowest purchasing power?''), whose gold answer is the country with the \emph{highest} inflation. Because participants in both arms overwhelmingly answered the question they were asked, we re-matched the gold and re-graded: the correction turned nine wrongly-failed submissions (two SQL-LLM, seven Snowflake) into passes and one previously-passing submission, which had answered the lowest-inflation reading, into a fail. This correction was made before the accuracy analyses were finalized; the completion-time model does not depend on grades, and the B-Q2 discrepancy below enters only as a sensitivity reading, never in the primary grades. The mismatch also drives the LLM-judge cross-check reported in the Measures (Section~\ref{subsec:measures}): the largest share of the judge-execution disagreements, 10 of 30, falls on M-Q2, where the judges had been supplied the same mismatched gold this audit corrects. The remaining disagreements scatter across the other tasks.

Ten of the remaining eleven golds pass the audit: the two \texttt{LIMIT 1} golds have unique optima, the most-prevalent-religion gold gives the same answer whether or not percentages are population-weighted, and the two with idiosyncratic output shapes (a \texttt{NULL} district row; a \texttt{SELECT *}) are graded by accepting any submission that returns the same underlying answer set. The remaining gold, B-Q2 (``How many authors are named Adam?''), is a prompt/gold discrepancy rather than mere ambiguity: the hint participants saw gives the literal predicate \texttt{author\_name LIKE 'Adam'}, which in SQL (absent wildcards) matches the exact string and so excludes ``Adam-Troy Castro,'' while the gold uses a prefix match that includes it. A participant who implemented the hint exactly as written is graded incorrect under the primary grading. Six participants answered under the hint's literal reading (one SQL-LLM, five Snowflake); re-grading those six as correct widens the gap to 46.7\% vs.\ 68.3\% and brings it to the significance boundary (exact per-participant permutation test, $p = 0.050$). Under either reading SQL-LLM confers no accuracy advantage.

\section{Per-Query Tests and Statistical Power}
\label{app:perquery}

Table~\ref{tab:ttest} reports the per-query two-sample $t$-tests with effect sizes and confidence intervals. These tests are exploratory: at $n = 10$ per arm each is underpowered, they are not adjusted for multiplicity, and no individual task carries inferential weight. Their role is descriptive, showing that the aggregate Group effect is consistent in direction across tasks (11 of 12 point estimates favor SQL-LLM) rather than resting on one or two queries. The inferential analysis is the mixed-effects model with its small-sample robustness block (Table~\ref{tab:overall}).

\begin{table}[t]
    \centering
    \begin{tabular}{lccc}
        \toprule
        \textbf{Query} & \textbf{t-statistic} & \textbf{p-value} & \textbf{Cohen's $d$ (95\% CI)} \\
        \midrule
        B-Q1 & $-1.41$ & $0.191$ & $-0.63$ $[-1.52, 0.28]$ \\
        B-Q2 & $\phantom{-}0.09$ & $0.927$ & $+0.04$ $[-0.84, 0.92]$ \\
        B-Q3 & $-2.26$ & $0.038$ & $-1.01$ $[-1.93, -0.06]$ \\
        B-Q4 & $-2.10$ & $0.060$ & $-0.94$ $[-1.85, 0.00]$ \\
        M-Q1 & $-1.80$ & $0.102$ & $-0.86$ $[-1.79, 0.10]$ \\
        M-Q2 & $-1.33$ & $0.211$ & $-0.64$ $[-1.55, 0.30]$ \\
        M-Q3 & $-0.36$ & $0.728$ & $-0.16$ $[-1.04, 0.72]$ \\
        M-Q4 & $-1.31$ & $0.209$ & $-0.58$ $[-1.47, 0.32]$ \\
        L-Q1 & $-0.81$ & $0.428$ & $-0.36$ $[-1.24, 0.53]$ \\
        L-Q2 & $-0.90$ & $0.381$ & $-0.40$ $[-1.29, 0.49]$ \\
        L-Q3 & $-1.45$ & $0.164$ & $-0.65$ $[-1.54, 0.26]$ \\
        L-Q4 & $-1.23$ & $0.246$ & $-0.55$ $[-1.44, 0.35]$ \\
        \bottomrule
    \end{tabular}
    \captionsetup{skip=5pt}
    \caption{Exploratory per-query two-sample $t$-tests on completion time, with between-subjects Cohen's $d$ and 95\% confidence intervals from the noncentral-$t$ distribution, the exact small-sample interval for a standardized difference (negative $d$ favors SQL-LLM); the per-query test and effect-size tables are included in the replication archive. The tests are unadjusted for multiplicity and individually underpowered ($n = 10$ per arm); read the effect estimates and intervals, not isolated $p$-values.}
    \label{tab:ttest}
\end{table}

\paragraph{Statistical power.}
With 10 participants per arm in a between-subjects design, the study is sized to detect medium-to-large interface effects. The observed effect on per-participant mean completion times is large: the raw between-arm difference is a 211 s reduction, matching the model's adjusted Group estimate of $\hat{\beta} = -212$ s to within rounding, or Cohen's $d \approx 0.89$. Computed at that observed effect size ($d \approx 0.89$, per-participant means, 10 per arm, two-sided $\alpha = 0.05$), a two-sample test achieves only about $47\%$ power, and reaching $80\%$ power at the same observed $d$ would require roughly $21$ participants per arm; a conventional medium effect ($d = 0.5$) would require about $64$ per arm. The same low power shapes how to read the per-query tests in Table~\ref{tab:ttest}: at $n = 10$ per arm, a per-query effect near the aggregate would be detected less than half the time, so the near-boundary per-query $p$-values are consistent with limited power and, on their own, neither establish nor rule out an effect at the task level.

A caveat on what the repeated measures buy: each participant contributes 12 observations, which sharpens the estimate of within-participant variation and lets the model partition individual baseline speed ($\sigma_{\text{participant}} \approx 201$ s) from the Group difference, but the repeated tasks do not create additional between-participant information. The Group effect is identified from the 20 participants to whom treatment was assigned, which is why the inference in Table~\ref{tab:overall} uses participant-level degrees of freedom and permutation checks rather than treating the 238 observations as independent evidence about the interface contrast.

\section{LLM-Coder Cross-Check}
\label{app:llmcoders}

In an earlier pass before the two-analyst coding, we ran three independent LLM coders (OpenAI GPT-5.5, Google Gemini~3~Pro Preview, and Anthropic Claude Opus~4.7) at \texttt{temperature=0} where supported, each asked to return a list of events (category, 5--25-word verbatim quote, one-sentence justification) rather than aggregate counts. Counts are derived as the size of each event list. Per-utterance annotations from all three coders are included in the replication archive (\texttt{llm\_annotations.jsonl}) so that any specific count can be audited.

Aggregate agreement between the LLM coders and that earlier single-coder pass (Pearson correlation over the 11 sessions) was strongest for the affective and exploratory categories, where frustration reached $r = 0.82$ and schema-check $r = 0.84$. The LLMs' absolute counts diverged unevenly rather than uniformly inflating (they over-counted most categories but under-counted hesitation), consistent with their varying in how strictly they applied the one-event-per-activity rule. These correlations predate the two-analyst coding in Section~\ref{sec:qualitative} and are retained in the replication archive (\texttt{coding\_agreement.csv}) for completeness.

\section{Changes from the First Preprint Version}
\label{app:preprint}

The first preprint version of this work (arXiv:2511.14718v1) reported results that this paper does not. In December 2025 we decided to revisit every analysis in the paper and build a replication package from which each reported result can be regenerated from the raw study data (Section~\ref{subsec:measures}). Anything we could not verify against the raw data at our disposal was dropped. The material changes fall into three groups.

\begin{enumerate}
    \item \textbf{Survey results.} The preprint described a post-task NASA-TLX and trust questionnaire and reported workload and trust comparisons between the arms. We found no reliable record that a structured instrument was properly administered, and no data from one; the only post-task material we hold is an open-ended debrief (Section~\ref{subsec:procedure}). The workload and trust results have been removed.
    \item \textbf{Accuracy and behavioral figures.} We could not trace the preprint's accuracy, frustration, reformulation, and error-recovery figures to graded submissions or to a reproducible analysis. This paper grades every submitted query by executing it against the BIRD gold databases, after the gold-answer audit in Appendix~\ref{app:goldaudit}, and draws its behavioral evidence from two independent coders (Section~\ref{sec:qualitative}). On this basis the accuracy comparison points the other way: 46\% for SQL-LLM versus 64\% for Snowflake ($p = 0.08$).
    \item \textbf{Setup.} The preprint stated that the study databases were hosted on PostgreSQL and that sessions ran in a laboratory on identical hardware. Both arms ran on Snowflake (Section~\ref{sec:methodology}), and sessions were remote and self-paced; the preprint's description reflected our own misunderstanding of the setup.
\end{enumerate}

The preprint had no replication package. This paper comes with one (see Data Availability), and every number reported here can be regenerated from it.

\bibliographystyle{ACM-Reference-Format}
\bibliography{references}


\begin{thebibliography}{32}


\ifx \showCODEN    \undefined \def \showCODEN     #1{\unskip}     \fi
\ifx \showISBNx    \undefined \def \showISBNx     #1{\unskip}     \fi
\ifx \showISBNxiii \undefined \def \showISBNxiii  #1{\unskip}     \fi
\ifx \showISSN     \undefined \def \showISSN      #1{\unskip}     \fi
\ifx \showLCCN     \undefined \def \showLCCN      #1{\unskip}     \fi
\ifx \shownote     \undefined \def \shownote      #1{#1}          \fi
\ifx \showarticletitle \undefined \def \showarticletitle #1{#1}   \fi
\ifx \showURL      \undefined \def \showURL       {\relax}        \fi
\providecommand\bibfield[2]{#2}
\providecommand\bibinfo[2]{#2}
\providecommand\natexlab[1]{#1}
\providecommand\showeprint[2][]{arXiv:#2}

\bibitem[Brass and Goldberg(2006)]%
        {brass2006semantic}
\bibfield{author}{\bibinfo{person}{Stefan Brass} {and}
  \bibinfo{person}{Christian Goldberg}.} \bibinfo{year}{2006}\natexlab{}.
\newblock \showarticletitle{Semantic errors in {SQL} queries: A quite complete
  list}.
\newblock \bibinfo{journal}{\emph{Journal of Systems and Software}}
  \bibinfo{volume}{79}, \bibinfo{number}{5} (\bibinfo{year}{2006}),
  \bibinfo{pages}{630--644}.
\newblock
\href{https://doi.org/10.1016/j.jss.2005.06.028}{doi:\nolinkurl{10.1016/j.jss.2005.06.028}}


\bibitem[Davis(1989)]%
        {davis1989tam}
\bibfield{author}{\bibinfo{person}{Fred~D. Davis}.}
  \bibinfo{year}{1989}\natexlab{}.
\newblock \showarticletitle{Perceived Usefulness, Perceived Ease of Use, and
  User Acceptance of Information Technology}.
\newblock \bibinfo{journal}{\emph{MIS Quarterly}} \bibinfo{volume}{13},
  \bibinfo{number}{3} (\bibinfo{year}{1989}), \bibinfo{pages}{319--340}.
\newblock
\href{https://doi.org/10.2307/249008}{doi:\nolinkurl{10.2307/249008}}


\bibitem[G{\"u}r et~al\mbox{.}(2018)]%
        {gur2018dialsql}
\bibfield{author}{\bibinfo{person}{Izzeddin G{\"u}r}, \bibinfo{person}{Semih
  Yavuz}, \bibinfo{person}{Yu Su}, {and} \bibinfo{person}{Xifeng Yan}.}
  \bibinfo{year}{2018}\natexlab{}.
\newblock \showarticletitle{{DialSQL}: Dialogue based Structured Query
  Generation}. In \bibinfo{booktitle}{\emph{Proceedings of the 56th Annual
  Meeting of the Association for Computational Linguistics (Volume 1: Long
  Papers)}}. \bibinfo{pages}{1339--1349}.
\newblock
\href{https://doi.org/10.18653/v1/P18-1124}{doi:\nolinkurl{10.18653/v1/P18-1124}}


\bibitem[Hart and Staveland(1988)]%
        {hart1988}
\bibfield{author}{\bibinfo{person}{S.~G. Hart} {and} \bibinfo{person}{L.~E.
  Staveland}.} \bibinfo{year}{1988}\natexlab{}.
\newblock \showarticletitle{Development of NASA-TLX (Task Load Index): Results
  of Empirical and Theoretical Research}.
\newblock In \bibinfo{booktitle}{\emph{Advances in Psychology}}.
  Vol.~\bibinfo{volume}{52}. \bibinfo{publisher}{Elsevier},
  \bibinfo{pages}{139--183}.
\newblock
\href{https://doi.org/10.1016/S0166-4115(08)62386-9}{doi:\nolinkurl{10.1016/S0166-4115(08)62386-9}}


\bibitem[Hoff and Bashir(2015)]%
        {hoff2015trust}
\bibfield{author}{\bibinfo{person}{Kevin~Anthony Hoff} {and}
  \bibinfo{person}{Masooda Bashir}.} \bibinfo{year}{2015}\natexlab{}.
\newblock \showarticletitle{Trust in Automation: Integrating Empirical Evidence
  on Factors That Influence Trust}.
\newblock \bibinfo{journal}{\emph{Human Factors}} \bibinfo{volume}{57},
  \bibinfo{number}{3} (\bibinfo{year}{2015}), \bibinfo{pages}{407--434}.
\newblock
\href{https://doi.org/10.1177/0018720814547570}{doi:\nolinkurl{10.1177/0018720814547570}}


\bibitem[Hong et~al\mbox{.}(2025)]%
        {hong2025next}
\bibfield{author}{\bibinfo{person}{Zijin Hong}, \bibinfo{person}{Zheng Yuan},
  \bibinfo{person}{Qinggang Zhang}, \bibinfo{person}{Hao Chen},
  \bibinfo{person}{Junnan Dong}, \bibinfo{person}{Feiran Huang}, {and}
  \bibinfo{person}{Xiao Huang}.} \bibinfo{year}{2025}\natexlab{}.
\newblock \showarticletitle{Next-Generation Database Interfaces: A Survey of
  {LLM}-Based Text-to-{SQL}}.
\newblock \bibinfo{journal}{\emph{IEEE Transactions on Knowledge and Data
  Engineering}} \bibinfo{volume}{37}, \bibinfo{number}{12}
  (\bibinfo{year}{2025}), \bibinfo{pages}{7328--7345}.
\newblock
\href{https://doi.org/10.1109/TKDE.2025.3609486}{doi:\nolinkurl{10.1109/TKDE.2025.3609486}}


\bibitem[Ioannidis and Simitsis(2009)]%
        {Simitsis2009}
\bibfield{author}{\bibinfo{person}{Yannis~E. Ioannidis} {and}
  \bibinfo{person}{Alkis Simitsis}.} \bibinfo{year}{2009}\natexlab{}.
\newblock \showarticletitle{{DBMSs} Should Talk Back Too}. In
  \bibinfo{booktitle}{\emph{Fourth Biennial Conference on Innovative Data
  Systems Research ({CIDR} 2009), Asilomar, CA, USA, January 4--7, 2009, Online
  Proceedings}}. \bibinfo{publisher}{www.cidrdb.org}.
\newblock
\showeprint[arxiv]{0909.1786}
\urldef\tempurl%
\url{https://arxiv.org/abs/0909.1786}
\showURL{%
\tempurl}


\bibitem[Jagadish et~al\mbox{.}(2007)]%
        {Jagadish2007}
\bibfield{author}{\bibinfo{person}{H.~V. Jagadish}, \bibinfo{person}{Adriane
  Chapman}, \bibinfo{person}{Aaron Elkiss}, \bibinfo{person}{Magesh
  Jayapandian}, \bibinfo{person}{Yunyao Li}, \bibinfo{person}{Arnab Nandi},
  {and} \bibinfo{person}{Cong Yu}.} \bibinfo{year}{2007}\natexlab{}.
\newblock \showarticletitle{Making database systems usable}. In
  \bibinfo{booktitle}{\emph{Proceedings of the 2007 ACM SIGMOD International
  Conference on Management of Data}} (Beijing, China)
  \emph{(\bibinfo{series}{SIGMOD '07})}. \bibinfo{publisher}{Association for
  Computing Machinery}, \bibinfo{address}{New York, NY, USA},
  \bibinfo{pages}{13–24}.
\newblock
\showISBNx{9781595936868}
\href{https://doi.org/10.1145/1247480.1247483}{doi:\nolinkurl{10.1145/1247480.1247483}}


\bibitem[Jarke et~al\mbox{.}(1985)]%
        {Jarke1985}
\bibfield{author}{\bibinfo{person}{Matthias Jarke}, \bibinfo{person}{Jon~A.
  Turner}, \bibinfo{person}{Edward~A. Stohr}, \bibinfo{person}{Yannis
  Vassiliou}, \bibinfo{person}{Norman~H. White}, {and} \bibinfo{person}{Ken
  Michielsen}.} \bibinfo{year}{1985}\natexlab{}.
\newblock \showarticletitle{A Field Evaluation of Natural Language for Data
  Retrieval}.
\newblock \bibinfo{journal}{\emph{IEEE Transactions on Software Engineering}}
  \bibinfo{volume}{SE-11}, \bibinfo{number}{1} (\bibinfo{year}{1985}),
  \bibinfo{pages}{97--114}.
\newblock
\href{https://doi.org/10.1109/TSE.1985.231847}{doi:\nolinkurl{10.1109/TSE.1985.231847}}


\bibitem[Kochedykov et~al\mbox{.}(2023)]%
        {kochedykov-etal-2023-conversing}
\bibfield{author}{\bibinfo{person}{Denis Kochedykov}, \bibinfo{person}{Fenglin
  Yin}, {and} \bibinfo{person}{Sreevidya Khatravath}.}
  \bibinfo{year}{2023}\natexlab{}.
\newblock \showarticletitle{Conversing with databases: Practical Natural
  Language Querying}. In \bibinfo{booktitle}{\emph{Proceedings of the 2023
  Conference on Empirical Methods in Natural Language Processing: Industry
  Track}}, \bibfield{editor}{\bibinfo{person}{Mingxuan Wang} {and}
  \bibinfo{person}{Imed Zitouni}} (Eds.). \bibinfo{publisher}{Association for
  Computational Linguistics}, \bibinfo{address}{Singapore},
  \bibinfo{pages}{372--379}.
\newblock
\href{https://doi.org/10.18653/v1/2023.emnlp-industry.36}{doi:\nolinkurl{10.18653/v1/2023.emnlp-industry.36}}


\bibitem[Krippendorff(2019)]%
        {krippendorff2004content}
\bibfield{author}{\bibinfo{person}{Klaus Krippendorff}.}
  \bibinfo{year}{2019}\natexlab{}.
\newblock \bibinfo{booktitle}{\emph{Content Analysis: An Introduction to Its
  Methodology} (\bibinfo{edition}{4th} ed.)}.
\newblock \bibinfo{publisher}{SAGE Publications}, \bibinfo{address}{Thousand
  Oaks, CA}.
\newblock
\href{https://doi.org/10.4135/9781071878781}{doi:\nolinkurl{10.4135/9781071878781}}


\bibitem[Lee and See(2004)]%
        {lee2004trust}
\bibfield{author}{\bibinfo{person}{John~D. Lee} {and}
  \bibinfo{person}{Katrina~A. See}.} \bibinfo{year}{2004}\natexlab{}.
\newblock \showarticletitle{Trust in Automation: Designing for Appropriate
  Reliance}.
\newblock \bibinfo{journal}{\emph{Human Factors}} \bibinfo{volume}{46},
  \bibinfo{number}{1} (\bibinfo{year}{2004}), \bibinfo{pages}{50--80}.
\newblock
\href{https://doi.org/10.1518/hfes.46.1.50_30392}{doi:\nolinkurl{10.1518/hfes.46.1.50_30392}}


\bibitem[Leventidis et~al\mbox{.}(2020)]%
        {Leventidis2020}
\bibfield{author}{\bibinfo{person}{Aristotelis Leventidis},
  \bibinfo{person}{Jiahui Zhang}, \bibinfo{person}{Cody Dunne},
  \bibinfo{person}{Wolfgang Gatterbauer}, \bibinfo{person}{H.~V. Jagadish},
  {and} \bibinfo{person}{Mirek Riedewald}.} \bibinfo{year}{2020}\natexlab{}.
\newblock \showarticletitle{QueryVis: Logic-based diagrams help users
  understand complicated SQL queries faster}. In
  \bibinfo{booktitle}{\emph{Proceedings of the 2020 ACM SIGMOD International
  Conference on Management of Data}}. \bibinfo{pages}{2303--2318}.
\newblock
\href{https://doi.org/10.1145/3318464.3389767}{doi:\nolinkurl{10.1145/3318464.3389767}}


\bibitem[Li and Jagadish(2014)]%
        {Li2014}
\bibfield{author}{\bibinfo{person}{Fei Li} {and} \bibinfo{person}{H.~V.
  Jagadish}.} \bibinfo{year}{2014}\natexlab{}.
\newblock \showarticletitle{Constructing an interactive natural language
  interface for relational databases}.
\newblock \bibinfo{journal}{\emph{Proceedings of the VLDB Endowment}}
  \bibinfo{volume}{8}, \bibinfo{number}{1} (\bibinfo{year}{2014}),
  \bibinfo{pages}{73--84}.
\newblock
\href{https://doi.org/10.14778/2735461.2735468}{doi:\nolinkurl{10.14778/2735461.2735468}}


\bibitem[Li and Jagadish(2016)]%
        {Li2016}
\bibfield{author}{\bibinfo{person}{Fei Li} {and} \bibinfo{person}{H.~V.
  Jagadish}.} \bibinfo{year}{2016}\natexlab{}.
\newblock \showarticletitle{Understanding Natural Language Queries over
  Relational Databases}.
\newblock \bibinfo{journal}{\emph{SIGMOD Record}} \bibinfo{volume}{45},
  \bibinfo{number}{1} (\bibinfo{year}{2016}), \bibinfo{pages}{6--13}.
\newblock
\href{https://doi.org/10.1145/2949741.2949744}{doi:\nolinkurl{10.1145/2949741.2949744}}


\bibitem[Li et~al\mbox{.}(2023)]%
        {li2024can}
\bibfield{author}{\bibinfo{person}{Jinyang Li}, \bibinfo{person}{Binyuan Hui},
  \bibinfo{person}{Ge Qu}, \bibinfo{person}{Jiaxi Yang},
  \bibinfo{person}{Binhua Li}, \bibinfo{person}{Bowen Li},
  \bibinfo{person}{Bailin Wang}, \bibinfo{person}{Bowen Qin},
  \bibinfo{person}{Ruiying Geng}, \bibinfo{person}{Nan Huo}, {et~al\mbox{.}}}
  \bibinfo{year}{2023}\natexlab{}.
\newblock \showarticletitle{Can {LLM} already serve as a database interface? a
  big bench for large-scale database grounded text-to-sqls}. In
  \bibinfo{booktitle}{\emph{Advances in Neural Information Processing Systems
  36 (NeurIPS 2023), Datasets and Benchmarks Track}}.
  \bibinfo{publisher}{Neural Information Processing Systems Foundation, Inc.},
  \bibinfo{pages}{42330--42357}.
\newblock
\href{https://doi.org/10.52202/075280-1835}{doi:\nolinkurl{10.52202/075280-1835}}


\bibitem[Luoma and Kumar(2025)]%
        {luoma2025snails}
\bibfield{author}{\bibinfo{person}{Kyle Luoma} {and} \bibinfo{person}{Arun
  Kumar}.} \bibinfo{year}{2025}\natexlab{}.
\newblock \showarticletitle{{SNAILS}: Schema Naming Assessments for Improved
  {LLM}-Based {SQL} Inference}.
\newblock \bibinfo{journal}{\emph{Proceedings of the ACM on Management of
  Data}} \bibinfo{volume}{3}, \bibinfo{number}{1}, Article
  \bibinfo{articleno}{77} (\bibinfo{year}{2025}).
\newblock
\href{https://doi.org/10.1145/3709727}{doi:\nolinkurl{10.1145/3709727}}


\bibitem[Miedema and Fletcher(2021)]%
        {Miedema2021}
\bibfield{author}{\bibinfo{person}{Daphne Miedema} {and}
  \bibinfo{person}{George H.~L. Fletcher}.} \bibinfo{year}{2021}\natexlab{}.
\newblock \showarticletitle{SQLVis: Visual Query Representations for Supporting
  SQL Learners}. In \bibinfo{booktitle}{\emph{2021 IEEE Symposium on Visual
  Languages and Human-Centric Computing (VL/HCC)}}. \bibinfo{pages}{1--9}.
\newblock
\href{https://doi.org/10.1109/VL/HCC51201.2021.9576431}{doi:\nolinkurl{10.1109/VL/HCC51201.2021.9576431}}


\bibitem[Mozannar et~al\mbox{.}(2024)]%
        {mozannar2024reading}
\bibfield{author}{\bibinfo{person}{Hussein Mozannar}, \bibinfo{person}{Gagan
  Bansal}, \bibinfo{person}{Adam Fourney}, {and} \bibinfo{person}{Eric
  Horvitz}.} \bibinfo{year}{2024}\natexlab{}.
\newblock \showarticletitle{Reading Between the Lines: Modeling User Behavior
  and Costs in {AI}-Assisted Programming}. In
  \bibinfo{booktitle}{\emph{Proceedings of the 2024 CHI Conference on Human
  Factors in Computing Systems}}. Article \bibinfo{articleno}{142}.
\newblock
\href{https://doi.org/10.1145/3613904.3641936}{doi:\nolinkurl{10.1145/3613904.3641936}}


\bibitem[Narechania et~al\mbox{.}(2021)]%
        {Narechania2021}
\bibfield{author}{\bibinfo{person}{Arpit Narechania}, \bibinfo{person}{Adam
  Fourney}, \bibinfo{person}{Bongshin Lee}, {and} \bibinfo{person}{Gonzalo
  Ramos}.} \bibinfo{year}{2021}\natexlab{}.
\newblock \showarticletitle{DIY: Assessing the correctness of natural language
  to SQL systems}. In \bibinfo{booktitle}{\emph{Proceedings of the 26th
  International Conference on Intelligent User Interfaces}}.
  \bibinfo{pages}{597--607}.
\newblock
\href{https://doi.org/10.1145/3397481.3450667}{doi:\nolinkurl{10.1145/3397481.3450667}}


\bibitem[Ning et~al\mbox{.}(2024)]%
        {ning2023empirical}
\bibfield{author}{\bibinfo{person}{Zheng Ning}, \bibinfo{person}{Yuan Tian},
  \bibinfo{person}{Zheng Zhang}, \bibinfo{person}{Tianyi Zhang}, {and}
  \bibinfo{person}{Toby Jia-Jun Li}.} \bibinfo{year}{2024}\natexlab{}.
\newblock \showarticletitle{Insights into Natural Language Database Query
  Errors: From Attention Misalignment to User Handling Strategies}.
\newblock \bibinfo{journal}{\emph{ACM Transactions on Interactive Intelligent
  Systems}} \bibinfo{volume}{14}, \bibinfo{number}{4}, Article
  \bibinfo{articleno}{25} (\bibinfo{year}{2024}), \bibinfo{numpages}{32}~pages.
\newblock
\showeprint[arxiv]{2402.07304}
\href{https://doi.org/10.1145/3650114}{doi:\nolinkurl{10.1145/3650114}}


\bibitem[Parasuraman and Riley(1997)]%
        {parasuraman1997humans}
\bibfield{author}{\bibinfo{person}{Raja Parasuraman} {and}
  \bibinfo{person}{Victor Riley}.} \bibinfo{year}{1997}\natexlab{}.
\newblock \showarticletitle{Humans and Automation: Use, Misuse, Disuse, Abuse}.
\newblock \bibinfo{journal}{\emph{Human Factors}} \bibinfo{volume}{39},
  \bibinfo{number}{2} (\bibinfo{year}{1997}), \bibinfo{pages}{230--253}.
\newblock
\href{https://doi.org/10.1518/001872097778543886}{doi:\nolinkurl{10.1518/001872097778543886}}


\bibitem[Pourreza and Rafiei(2023)]%
        {pourreza2023gpt4}
\bibfield{author}{\bibinfo{person}{Mohammadreza Pourreza} {and}
  \bibinfo{person}{Davood Rafiei}.} \bibinfo{year}{2023}\natexlab{}.
\newblock \showarticletitle{DIN-SQL: decomposed in-context learning of
  text-to-SQL with self-correction}. In \bibinfo{booktitle}{\emph{Proceedings
  of the 37th International Conference on Neural Information Processing
  Systems}} (New Orleans, LA, USA) \emph{(\bibinfo{series}{NIPS '23})}.
  \bibinfo{publisher}{Curran Associates Inc.}, \bibinfo{address}{Red Hook, NY,
  USA}, Article \bibinfo{articleno}{1577}, \bibinfo{numpages}{10}~pages.
\newblock
\href{https://doi.org/10.52202/075280-1577}{doi:\nolinkurl{10.52202/075280-1577}}


\bibitem[Shneiderman(1978)]%
        {Shneiderman1978}
\bibfield{author}{\bibinfo{person}{Ben Shneiderman}.}
  \bibinfo{year}{1978}\natexlab{}.
\newblock \showarticletitle{Improving the Human Factors Aspect of Database
  Interactions}.
\newblock \bibinfo{journal}{\emph{ACM Transactions on Database Systems}}
  \bibinfo{volume}{3}, \bibinfo{number}{4} (\bibinfo{year}{1978}),
  \bibinfo{pages}{417--439}.
\newblock
\href{https://doi.org/10.1145/320289.320295}{doi:\nolinkurl{10.1145/320289.320295}}


\bibitem[Taipalus et~al\mbox{.}(2018)]%
        {taipalus2018errors}
\bibfield{author}{\bibinfo{person}{Toni Taipalus}, \bibinfo{person}{Mikko
  Siponen}, {and} \bibinfo{person}{Tero Vartiainen}.}
  \bibinfo{year}{2018}\natexlab{}.
\newblock \showarticletitle{Errors and Complications in {SQL} Query
  Formulation}.
\newblock \bibinfo{journal}{\emph{ACM Transactions on Computing Education}}
  \bibinfo{volume}{18}, \bibinfo{number}{3}, Article \bibinfo{articleno}{15}
  (\bibinfo{year}{2018}), \bibinfo{numpages}{29}~pages.
\newblock
\href{https://doi.org/10.1145/3231712}{doi:\nolinkurl{10.1145/3231712}}


\bibitem[Tian et~al\mbox{.}(2024)]%
        {tian2024sqlucid}
\bibfield{author}{\bibinfo{person}{Yuan Tian}, \bibinfo{person}{Jonathan~K.
  Kummerfeld}, \bibinfo{person}{Toby Jia-Jun Li}, {and} \bibinfo{person}{Tianyi
  Zhang}.} \bibinfo{year}{2024}\natexlab{}.
\newblock \showarticletitle{SQLucid: Grounding Natural Language Database
  Queries with Interactive Explanations}. In
  \bibinfo{booktitle}{\emph{Proceedings of the 37th Annual ACM Symposium on
  User Interface Software and Technology}} (Pittsburgh, PA, USA)
  \emph{(\bibinfo{series}{UIST '24})}. \bibinfo{publisher}{Association for
  Computing Machinery}, \bibinfo{address}{New York, NY, USA}, Article
  \bibinfo{articleno}{12}, \bibinfo{numpages}{20}~pages.
\newblock
\showISBNx{9798400706288}
\href{https://doi.org/10.1145/3654777.3676368}{doi:\nolinkurl{10.1145/3654777.3676368}}


\bibitem[Tian et~al\mbox{.}(2023)]%
        {tian2023steps}
\bibfield{author}{\bibinfo{person}{Yuan Tian}, \bibinfo{person}{Zheng Zhang},
  \bibinfo{person}{Zheng Ning}, \bibinfo{person}{Toby Jia-Jun Li},
  \bibinfo{person}{Jonathan~K. Kummerfeld}, {and} \bibinfo{person}{Tianyi
  Zhang}.} \bibinfo{year}{2023}\natexlab{}.
\newblock \showarticletitle{Interactive Text-to-{SQL} Generation via Editable
  Step-by-Step Explanations}. In \bibinfo{booktitle}{\emph{Proceedings of the
  2023 Conference on Empirical Methods in Natural Language Processing}}.
  \bibinfo{publisher}{Association for Computational Linguistics},
  \bibinfo{address}{Singapore}, \bibinfo{pages}{16149--16166}.
\newblock
\href{https://doi.org/10.18653/v1/2023.emnlp-main.1004}{doi:\nolinkurl{10.18653/v1/2023.emnlp-main.1004}}


\bibitem[Vassiliou et~al\mbox{.}(1983)]%
        {Vassiliou1983}
\bibfield{author}{\bibinfo{person}{Yannis Vassiliou}, \bibinfo{person}{Matthias
  Jarke}, \bibinfo{person}{Edward~A. Stohr}, \bibinfo{person}{Jon~A. Turner},
  {and} \bibinfo{person}{Norman~H. White}.} \bibinfo{year}{1983}\natexlab{}.
\newblock \showarticletitle{Natural Language for Database Queries: A Laboratory
  Study}.
\newblock \bibinfo{journal}{\emph{MIS Quarterly}} \bibinfo{volume}{7},
  \bibinfo{number}{4} (\bibinfo{year}{1983}), \bibinfo{pages}{47--61}.
\newblock
\href{https://doi.org/10.2307/248746}{doi:\nolinkurl{10.2307/248746}}


\bibitem[Venkatesh et~al\mbox{.}(2003)]%
        {venkatesh2003utaut}
\bibfield{author}{\bibinfo{person}{Viswanath Venkatesh},
  \bibinfo{person}{Michael~G. Morris}, \bibinfo{person}{Gordon~B. Davis}, {and}
  \bibinfo{person}{Fred~D. Davis}.} \bibinfo{year}{2003}\natexlab{}.
\newblock \showarticletitle{User Acceptance of Information Technology: Toward a
  Unified View}.
\newblock \bibinfo{journal}{\emph{MIS Quarterly}} \bibinfo{volume}{27},
  \bibinfo{number}{3} (\bibinfo{year}{2003}), \bibinfo{pages}{425--478}.
\newblock
\href{https://doi.org/10.2307/30036540}{doi:\nolinkurl{10.2307/30036540}}


\bibitem[Wretblad et~al\mbox{.}(2024)]%
        {wretblad2024understanding}
\bibfield{author}{\bibinfo{person}{Niklas Wretblad}, \bibinfo{person}{Fredrik
  Riseby}, \bibinfo{person}{Rahul Biswas}, \bibinfo{person}{Amin Ahmadi}, {and}
  \bibinfo{person}{Oskar Holmstr{\"o}m}.} \bibinfo{year}{2024}\natexlab{}.
\newblock \showarticletitle{Understanding the Effects of Noise in
  Text-to-{SQL}: An Examination of the {BIRD}-Bench Benchmark}. In
  \bibinfo{booktitle}{\emph{Proceedings of the 62nd Annual Meeting of the
  Association for Computational Linguistics (Volume 2: Short Papers)}}.
  \bibinfo{pages}{356--369}.
\newblock
\href{https://doi.org/10.18653/v1/2024.acl-short.34}{doi:\nolinkurl{10.18653/v1/2024.acl-short.34}}


\bibitem[Yao et~al\mbox{.}(2019)]%
        {Yao2019}
\bibfield{author}{\bibinfo{person}{Ziyu Yao}, \bibinfo{person}{Yu Su},
  \bibinfo{person}{Huan Sun}, {and} \bibinfo{person}{Wen-tau Yih}.}
  \bibinfo{year}{2019}\natexlab{}.
\newblock \showarticletitle{Model-Based Interactive Semantic Parsing: A Unified
  Framework and a Text-to-SQL Case Study}. In
  \bibinfo{booktitle}{\emph{Proceedings of the 2019 Conference on Empirical
  Methods in Natural Language Processing and the 9th International Joint
  Conference on Natural Language Processing (EMNLP-IJCNLP)}}.
  \bibinfo{pages}{5447--5458}.
\newblock
\href{https://doi.org/10.18653/v1/D19-1547}{doi:\nolinkurl{10.18653/v1/D19-1547}}


\bibitem[Yu et~al\mbox{.}(2018)]%
        {yu2018spider}
\bibfield{author}{\bibinfo{person}{Tao Yu}, \bibinfo{person}{Rui Zhang},
  \bibinfo{person}{Kai Yang}, \bibinfo{person}{Michihiro Yasunaga},
  \bibinfo{person}{Dongxu Wang}, \bibinfo{person}{Zifan Li},
  \bibinfo{person}{James Ma}, \bibinfo{person}{Irene Li},
  \bibinfo{person}{Qingning Yao}, \bibinfo{person}{Shanelle Roman},
  \bibinfo{person}{Zilin Zhang}, {and} \bibinfo{person}{Dragomir Radev}.}
  \bibinfo{year}{2018}\natexlab{}.
\newblock \showarticletitle{Spider: A Large-Scale Human-Labeled Dataset for
  Complex and Cross-Domain Semantic Parsing and Text-to-SQL Task}. In
  \bibinfo{booktitle}{\emph{Proceedings of the 2018 Conference on Empirical
  Methods in Natural Language Processing}}. \bibinfo{pages}{3911--3921}.
\newblock
\href{https://doi.org/10.18653/v1/D18-1425}{doi:\nolinkurl{10.18653/v1/D18-1425}}


\end{thebibliography}

\end{document}